\documentclass[11pt]{article}
\usepackage{tikz}
\usepackage{geometry}
\usepackage[english]{babel}
\usepackage[utf8]{inputenc}
\usepackage[T1]{fontenc}
\usepackage{indentfirst}
\usepackage{amsmath}
\usepackage{amssymb}
\usepackage{amsthm}
\usepackage{proof}
\usepackage{eufrak}
\usepackage{graphicx}
\usepackage{psfrag}

\allowhyphens

\newcommand{\ket}[1]{{\left|#1\right\rangle}}

\pgfdeclarelayer{edgelayer}
\pgfdeclarelayer{nodelayer}
\pgfsetlayers{edgelayer,nodelayer,main}
\tikzstyle{none}=[inner sep=0pt]
\tikzstyle{rn}=[circle,fill=Red,draw=Black,line width=0.8 pt]
\tikzstyle{gn}=[circle,fill=Lime,draw=Black,line width=0.8 pt]
\tikzstyle{yn}=[circle,fill=Yellow,draw=Black,line width=0.8 pt]
\tikzstyle{simple}=[-,draw=Black,line width=2.000]
\tikzstyle{arrow}=[-,draw=Black,postaction={decorate},decoration={markings,mark=at position .5 with {\arrow{>}}},line width=2.000]
\tikzstyle{tick}=[-,draw=Black,postaction={decorate},decoration={markings,mark=at position .5 with {\draw (0,-0.1) -- (0,0.1);}},line width=2.000]
\usetikzlibrary{arrows}
\tikzstyle{black}=[fill=black, shape=circle]
\tikzstyle{arrow}=[->]
\tikzstyle{red arrow}=[draw=red, ->]


\newcommand{\be}{\begin{equation}}
\newcommand{\ee}{\end{equation}}
\newcommand{\bea}{\begin{eqnarray}}
\newcommand{\eea}{\end{eqnarray}}

\usepackage{ifpdf}

\ifpdf
\usepackage{epstopdf}
\usepackage[pdftex,colorlinks,urlcolor=blue,citecolor=blue,linkcolor=black]{hyperref}
\else
\usepackage[hypertex,colorlinks,urlcolor=blue,citecolor=blue,linkcolor=black]{hyperref}
\fi
\begin{document}

\thispagestyle{empty}

\begin{center}%
{\LARGE\textbf{\mathversion{bold}%
Temperley-Lieb integrable models and fusion categories}\par}

\vspace{1cm}
{\textsc{Matthew Blakeney$^{a, b, c}$, Luke Corcoran$^{ a}$, Marius de Leeuw$^{ a}$, Bal{\'a}zs Pozsgay$^{ d}$, Eric Vernier$^{ e}$ }}
\vspace{8mm} \\
\textit{
$^a$ School of Mathematics \& Hamilton Mathematics Institute, \\
Trinity College Dublin, Ireland\\
[5pt]
}
\vspace{.2cm}
\textit{
$^b$  Perimeter Institute for Theoretical Physics, Waterloo, Canada \\
[5pt]
}
\vspace{.2cm}
\textit{
$^c$  Department of Physics and Astronomy,\\
University of Waterloo, Waterloo, ON, Canada\\
[5pt]
}
\vspace{.2cm}
\textit{
$^d$ MTA-ELTE “Momentum” Integrable Quantum Dynamics Research Group, \\
Department of Theoretical Physics,\\
ELTE Eötvös Loránd University, Budapest, Hungary\\
[5pt]
}
\vspace{.2cm}
\textit{
$^e$ Laboratoire de Probabilit\'es, Statistique et Mod\'elisation \& CNRS,
Universit\'e Paris Cit\'e, Sorbonne Universit\'e Paris, France \\[5pt]
}
\texttt{\{blakenem, corcorl2, deleeuwm\}@tcd.ie} \\
\texttt{ pozsgay.balazs@ttk.elte.hu} \\
\texttt{vernier@lpsm.paris}
%

\par\vspace{15mm}

\textbf{Abstract} \vspace{5mm}

\begin{minipage}{13cm}

We show that every fusion category containing a non-invertible, self-dual object $a$ gives rise to an integrable anyonic chain whose Hamiltonian density satisfies the Temperley-Lieb algebra. This spin chain arises by considering the projection onto the identity channel in the fusion process $a\otimes a$. We relate these models to Pasquier's construction of ADE lattice models. We then exploit the underlying Temperley-Lieb structure to discuss the spectrum of these models and argue that these models are gapped when the quantum dimension of $a$ is greater than 2. We show that for fusion categories where the dimension is close to 2, such as the Fib$\times$Fib and Haagerup fusion categories, the finite size effects are large and they can obscure the numerical analysis of the gap.
 
\end{minipage}
\end{center}

\newpage 

\tableofcontents
\bigskip
\hrule

\section{Introduction}
Anyonic spin chains have received a great deal of attention in recent years; partly due to their connections to conformal field theories (CFTs) and critical phenomena \cite{Buican:2017rxc}, and partly due to their realisation of non-invertible topological symmetries \cite{McGreevy:2022oyu,Cordova:2022ruw,Shao:2023gho,Schafer-Nameki:2023jdn,Bhardwaj:2023kri}. The simplest example is the golden chain \cite{Feiguin:2006ydp, Trebst_2008}, which describes the interaction of Fibonacci anyons and can be regarded as an anyonic analog of the Heisenberg XXX spin chain. The golden chain enjoys several interesting properties. Firstly, it is an integrable model whose Hamiltonian densities satisfy the Temperley-Lieb algebra \cite{TemperleyLieb1971, Wang_1996}. Secondly, the model is critical and tends to a CFT at both ends of the spectrum in the continuum limit. In particular, the ground state tends to that of the $c=7/10$ tricritical Ising model, and the anti-ground state tends to the ground state of the $c=4/5$ three-state Potts model. Finally, the golden chain has a topological symmetry which furnishes a representation of the Fibonacci fusion category.

Fusion categories play an important role in several areas of physics, including conformal field theory \cite{VERLINDE1988360} and topological quantum field theory \cite{Chang:2018iay}. Many reviews exist from both a mathematical \cite{Kirillov, etingof2017fusioncategories} and physical \cite{KITAEV20062} perspective. A systematic way of constructing 2d statistical physics models from fusion categories was proposed by Aasen, Fendley, and Mong \cite{Aasen:2016dop, Aasen:2020jwb}. In the anisotropic limit one obtains the anyonic chain Hamiltonian, which is defined by an input fusion category $\mathcal{C}$ and an object $a \in \mathcal{C}$. The object $a$ is the `external' object of the chain, and takes the role of the particle being fused. The anyonic chain Hamiltonian is then given as the projection of neighbouring $a$ objects onto an object $b \in a\otimes a$, or a linear combination of such projections. The golden chain is given by the case $\mathcal{C} = \text{Fib} = \{1, \tau \}$, with external object $a = \tau$ and fusion channel $b = 1$.

Several generalisations of the golden chain have been considered, and it is interesting to study the survival/absence of integrability and criticality as one moves away from this simplest case. The Fibonacci fusion category can be regarded as $\mathfrak{psu}(2)_k$, a projection of the truncated angular momentum fusion algebra, for $k = 4$. One way to go beyond the golden chain is to increase $k$, and consider external objects $a$ which give rise to `higher spin' anyonic chains \cite{Gils_2013, Vernier_2017}.\footnote{Here `spin' refers to the number of degrees of freedom at each site of the chain. We say a model is spin $s$ if $a \otimes a$ contains $2s + 1$ objects. In practice the number of spins at each site is less than this due to fusion constraints.} One can also break the topological symmetry while maintaining integrability, which leads to the off-critical golden chain \cite{Fendley_2004, Bianchini:2014bfa}. While there is no fusion category symmetry, the constraints from the fusion algebra remain, and the model is essentially one of Rydberg atoms \cite{Urban2008ObservationOR, PhysRevA.86.041601, Lesanovsky_2012, Turner_2018,  Corcoran:2024ofo}. Finally, one can consider anyonic chains built from fusion categories which are not related to a deformation of the $\mathfrak{su}(2)$ algebra. For example, taking $\mathcal{C} = \text{Rep}(D_3)$ gives a chain with symmetries based on representations of the dihedral group \cite{Braylovskaya:2016btd, Eck:2023gic}. The $F$-symbols of the Haagerup fusion category $\mathcal{H}_3$ \cite{haagerup, Grossman_2012} were calculated recently \cite{Huang:2020lox,Wolf:2020qdo,osborne2019fsymbols}, which allowed for an exploration into models with this more exotic symmetry \cite{Huang:2021nvb, Vanhove:2021zop,Corcoran:2024eeh}. Models based on products of Fibonacci categories have also been investigated recently  \cite{Antunes:2025huk}. The study of anyonic chains based on different fusion categories has been aided by recent efforts to catalogue categories of low rank, and calculate their $F$-symbols algorithmically \cite{gert_thesis, vercleyen_anyonica_2025, vercleyen_slingerland_anyonwiki_2025}.

In this paper we study anyonic chains based on general fusion categories $\mathcal{C}$, restricting to the identity fusion channel $b = 1$. We prove that for any choice of external object $a$ the model obeys the Temperley-Lieb algebra with parameter $\delta = \text{dim}\hspace{0.05cm}a$, the quantum dimension of $a$. As such, the models are integrable, a fact that we revisit by constructing the corresponding transfer matrices within the recently developed framework of medium-range integrability \cite{Gombor:2021nhn}. 
These models are related to the critical family of Temperley-Lieb models constructed by Pasquier from ADE type Dynkin diagrams \cite{Pasquier:1987xj,PASQUIER1987162}. Our models differ from these, however, as they allow for Temperley-Lieb parameters $\delta > 2$. In these cases we demonstrate the models are \textit{not} critical, and in fact gapped. There has been confusion in the literature about this issue, as the departure from criticality in these cases is often subtle \cite{Gils_2013, Corcoran:2024eeh, Antunes:2025huk}. The XXZ spin chain is a one-parameter family of integrable models which satisfies the Temperley-Lieb algebra with $\delta = 2\Delta$, and is critical when $-1 < \Delta \leq 1$ \cite{PhysRevB.34.6372, korepin1993quantuminversescatteringmethod}. Since the anyonic chains and the XXZ chain are both Temperley-Lieb, we are able to decompose the anyonic chain spectrum in terms of an XXZ model with $\Delta =  \text{dim}\hspace{0.05cm}a/2$ \cite{ Vernier_2017,Fukai_2024}. We can then explore the spectrum of the model to very high lengths using Bethe ansatz methods, and argue that they are gapped for $ \text{dim}\hspace{0.05cm}a =\delta> 2$. We do this explicitly for $\mathcal{C} = \mathcal{H}_3$ and $\mathcal{C} = \text{Fib}\times \text{Fib}$. Other cases such as $\mathcal{C} = \mathfrak{psu}(2)_k$ can be treated similarly, and this was done in the case of spin-1 external object in \cite{Vernier_2017}. We also note that similar results have been obtained in the case of string-net models \cite{Schulz:2014cfa}, particularly in the case of $\mathfrak{psu}(2)_5$ fusion rules.

The plan for the paper is as follows. In section \ref{sec:fusionandanyon} we give an introduction to fusion categories and anyonic chains, giving several examples. In section \ref{sec:TL} we prove the main result of our paper: given a fusion category $\mathcal{C}$ and an object $a$ such that $a\otimes a = 1\oplus\cdots$, the anyonic chain $p^{(a,1)}$ which projects $a\otimes a \rightarrow 1$ satisfies the Temperley-Lieb algebra. We relate these models to Pasquier's ADE lattice models and discuss their integrability. Finally in section \ref{sec:gap} we argue that in the case of unitary fusion categories and $\text{dim}\hspace{0.05cm} a > 2$, these TL models are gapped. We do this explicitly in the cases $\mathcal{C}=\text{Fib}\times \text{Fib}$ and $\mathcal{C}=\mathcal{H}_3$, where there has been confusion previously about the gapped nature of these chains.

\section{Fusion Categories and Anyonic Chains}\label{sec:fusionandanyon}

In this section we give a brief introduction to fusion categories and anyonic chains. We keep the discussion physically motivated and only introduce concepts that are relevant for the models we discuss in this paper. We refer to \cite{Kirillov, etingof2017fusioncategories} for mathematical introductions to fusion categories and \cite{Trebst_2008, Aasen:2020jwb} for physically motivated introductions. There are several available reviews of anyonic chains, see for example \cite{Shao:2023gho, Eck:2024myo}.

\subsection{Fusion rules} The particles/topological charges that constitute our models and the constraints on our Hilbert space are defined by fusion rules. Fusion rules are defined by a set of objects $B$, and a set of non-negative integers $N_{ab}^{ c}$ such that
\begin{equation}
a\otimes b =\bigoplus_{c\in B}N_{ab}^c \hspace{0.1cm} c,
\end{equation}
where $a$ and $b$ are the objects being fused and $c$ is the result of their fusion. The cardinality of $B$, $|B|$, is the rank of the fusion rules. Fusion rules which meet reasonable conditions for physicality define a fusion ring, with rank $|B|$. One such condition is that for any object $a$, there exists a unique dual object $\bar{a}$, such that $N_{a,\bar{b}}^1=\delta_{a,b}$. The models considered in this paper do not require symmetric fusion rules, because no braiding structure is required.

\par
The fusion of particles is represented graphically by the fusion of edges on a graph. Given neighbouring particles $a, b,$ and $c$, there are two ways to combine them into a single particle $d$:
\begin{equation*}
\begin{tikzpicture}[scale=0.7]
	\begin{pgfonlayer}{nodelayer}
		\node [style=none] (0) at (6.3125, -1.75) {and};
		\node [style=none] (1) at (12.125, -3.875) {.};
		\node [style=none] (2) at (0.5, 0) {};
		\node [style=none] (3) at (7.5, 0) {};
		\node [style=black] (4) at (2.5, 0) {};
		\node [style=black] (5) at (4.5, 0) {};
		\node [style=black] (6) at (6.5, 0) {};
		\node [style=none] (7) at (0.5, 0) {};
		\node [style=none] (8) at (7.5, 0) {};
		\node [style=none] (9) at (1.5, 0) {};
		\node [style=none] (10) at (5.5, 0) {};
		\node [style=none] (11) at (1.5, 0) {};
		\node [style=none] (12) at (5.5, 0) {};
		\node [style=none] (13) at (3.075, 0) {$a$};
		\node [style=none] (14) at (3.925, 0.075) {$b$};
		\node [style=none] (15) at (5.925, 0) {$c$};
		\node [style=none] (16) at (8.25, 0) {$=$};
		\node [style=none] (17) at (8.875, 1.25) {};
		\node [style=none] (18) at (10.375, 1.25) {};
		\node [style=none] (19) at (11.875, 1.25) {};
		\node [style=none] (20) at (10.375, -0.25) {};
		\node [style=none] (21) at (9.125, 1) {};
		\node [style=none] (22) at (10.125, 1) {};
		\node [style=none] (23) at (9.625, 0.5) {};
		\node [style=none] (24) at (9.875, 0.25) {};
		\node [style=none] (25) at (11.375, 0.75) {};
		\node [style=none] (26) at (10.375, -1.25) {};
		\node [style=none] (27) at (10.375, -1) {};
		\node [style=none] (28) at (8.25, -3.875) {$=$};
		\node [style=none] (29) at (11.875, -2.625) {};
		\node [style=none] (30) at (10.375, -2.625) {};
		\node [style=none] (31) at (8.875, -2.625) {};
		\node [style=none] (32) at (10.375, -4.125) {};
		\node [style=none] (33) at (11.625, -2.875) {};
		\node [style=none] (34) at (10.625, -2.875) {};
		\node [style=none] (35) at (11.125, -3.375) {};
		\node [style=none] (36) at (10.875, -3.625) {};
		\node [style=none] (37) at (9.375, -3.125) {};
		\node [style=none] (38) at (10.375, -5.125) {};
		\node [style=none] (39) at (10.375, -4.875) {};
		\node [style=none] (40) at (8.875, 1.625) {$a$};
		\node [style=none] (42) at (11.875, 1.625) {$c$};
		\node [style=none] (43) at (8.875, -2.25) {$a$};
		\node [style=none] (44) at (10.375, -2.175) {$b$};
		\node [style=none] (45) at (11.875, -2.25) {$c$};
		\node [style=none] (46) at (6.625, -1) {$d$};
		\node [style=none] (47) at (5.475, -0.55) {$e$};
		\node [style=none] (48) at (10.625, -4.875) {$d$};
		\node [style=none] (49) at (10.625, -1) {$d$};
		\node [style=none] (50) at (9.7, -0.025) {$e$};
		\node [style=none] (51) at (11.25, -3.9) {$f$};
		\node [style=none] (52) at (7.5, -3.875) {};
		\node [style=none] (53) at (0.5, -3.875) {};
		\node [style=black] (54) at (5.5, -3.875) {};
		\node [style=black] (55) at (3.5, -3.875) {};
		\node [style=black] (56) at (1.5, -3.875) {};
		\node [style=none] (57) at (7.5, -3.875) {};
		\node [style=none] (58) at (0.5, -3.875) {};
		\node [style=none] (59) at (6.5, -3.875) {};
		\node [style=none] (60) at (2.5, -3.875) {};
		\node [style=none] (61) at (6.5, -3.875) {};
		\node [style=none] (62) at (2.5, -3.875) {};
		\node [style=none] (63) at (2.075, -3.875) {$a$};
		\node [style=none] (64) at (4.075, -3.8) {$b$};
		\node [style=none] (65) at (4.9, -3.875) {$c$};
		\node [style=none] (66) at (1.25, -4.875) {$d$};
		\node [style=none] (67) at (2.525, -4.425) {$f$};
		\node [style=none] (68) at (12.125, 0) {,};
		\node [style=none] (69) at (8.875, 1.625) {$a$};
		\node [style=none] (70) at (10.375, 1.7) {$b$};
		\node [style=none] (71) at (11.875, 1.625) {$c$};
	\end{pgfonlayer}
	\begin{pgfonlayer}{edgelayer}
		\draw [in=90, out=90, looseness=0.50] (2.center) to (3.center);
		\draw [in=-90, out=-90, looseness=0.50] (7.center) to (8.center);
		\draw [in=-90, out=-90, looseness=0.50] (9.center) to (10.center);
		\draw [in=90, out=90, looseness=0.50] (11.center) to (12.center);
		\draw (21.center) to (23.center);
		\draw (23.center) to (22.center);
		\draw (24.center) to (20.center);
		\draw (20.center) to (25.center);
		\draw (27.center) to (26.center);
		\draw (33.center) to (35.center);
		\draw (35.center) to (34.center);
		\draw (36.center) to (32.center);
		\draw (32.center) to (37.center);
		\draw (39.center) to (38.center);
		\draw (17.center) to (21.center);
		\draw (22.center) to (18.center);
		\draw (25.center) to (19.center);
		\draw (20.center) to (27.center);
		\draw (24.center) to (23.center);
		\draw (37.center) to (31.center);
		\draw (34.center) to (30.center);
		\draw (33.center) to (29.center);
		\draw (36.center) to (35.center);
		\draw (39.center) to (32.center);
		\draw [in=90, out=90, looseness=0.50] (52.center) to (53.center);
		\draw [in=-90, out=-90, looseness=0.50] (57.center) to (58.center);
		\draw [in=-90, out=-90, looseness=0.50] (59.center) to (60.center);
		\draw [in=90, out=90, looseness=0.50] (61.center) to (62.center);
	\end{pgfonlayer}
\end{tikzpicture}
\end{equation*}
\subsection{Fusion categories} Physically, as there are two distinct procedures by which one may fuse three particles, change of basis matrices are required. These are called associators/$F$-symbols and in this paper the convention followed is
\begin{equation}
\begin{tikzpicture}[scale=0.7]
	\begin{pgfonlayer}{nodelayer}
		\node [style=none] (86) at (2, -3.5) {$=$};
		\node [style=none] (87) at (-1.75, -2.25) {};
		\node [style=none] (88) at (-0.25, -2.25) {};
		\node [style=none] (89) at (1.25, -2.25) {};
		\node [style=none] (90) at (-0.25, -3.75) {};
		\node [style=none] (91) at (-1.5, -2.5) {};
		\node [style=none] (92) at (-0.5, -2.5) {};
		\node [style=none] (93) at (-1, -3) {};
		\node [style=none] (94) at (-0.75, -3.25) {};
		\node [style=none] (95) at (0.75, -2.75) {};
		\node [style=none] (96) at (-0.25, -4.75) {};
		\node [style=none] (97) at (-0.25, -4.5) {};
		\node [style=none] (98) at (8.7, -2.25) {};
		\node [style=none] (99) at (7.2, -2.25) {};
		\node [style=none] (100) at (5.7, -2.25) {};
		\node [style=none] (101) at (7.2, -3.75) {};
		\node [style=none] (102) at (8.45, -2.5) {};
		\node [style=none] (103) at (7.45, -2.5) {};
		\node [style=none] (104) at (7.95, -3) {};
		\node [style=none] (105) at (7.7, -3.25) {};
		\node [style=none] (106) at (6.2, -2.75) {};
		\node [style=none] (107) at (7.2, -4.75) {};
		\node [style=none] (108) at (7.2, -4.5) {};
		\node [style=none] (109) at (-1.75, -1.875) {$a$};
		\node [style=none] (110) at (-0.25, -1.825) {$b$};
		\node [style=none] (111) at (1.25, -1.875) {$c$};
		\node [style=none] (112) at (5.7, -1.875) {$a$};
		\node [style=none] (113) at (7.2, -1.825) {$b$};
		\node [style=none] (114) at (8.7, -1.875) {$c$};
		\node [style=none] (115) at (7.45, -4.5) {$d$};
		\node [style=none] (116) at (0, -4.5) {$d$};
		\node [style=none] (117) at (-0.925, -3.525) {$e$};
		\node [style=none] (118) at (8.075, -3.525) {$f$};
		\node [style=none] (119) at (8.95, -3.55) {.};
		\node [style=none, align=left] (120) at (3.875, -3.6) {$\sum\limits_{f\in B}(F_d^{a,b,c})_{e,f}$};
	\end{pgfonlayer}
	\begin{pgfonlayer}{edgelayer}
		\draw (91.center) to (93.center);
		\draw (93.center) to (92.center);
		\draw (94.center) to (90.center);
		\draw (90.center) to (95.center);
		\draw (97.center) to (96.center);
		\draw (102.center) to (104.center);
		\draw (104.center) to (103.center);
		\draw (105.center) to (101.center);
		\draw (101.center) to (106.center);
		\draw (108.center) to (107.center);
		\draw (87.center) to (91.center);
		\draw (92.center) to (88.center);
		\draw (95.center) to (89.center);
		\draw (90.center) to (97.center);
		\draw (94.center) to (93.center);
		\draw (106.center) to (100.center);
		\draw (103.center) to (99.center);
		\draw (102.center) to (98.center);
		\draw (105.center) to (104.center);
		\draw (108.center) to (101.center);
	\end{pgfonlayer}
\end{tikzpicture} \label{eq:basischange}
\end{equation}
As we constrict our attention to the `multiplicity-free' case where $N_{ab}^c\in \{0,1\}$, no additional indices are required on the $F$-symbols. A constraint on the values of $F$-symbols is provided by the \textit{pentagon axiom}, which states that the two alternative compositions of associators which completely change the basis of the four-particle fusion diagram must be equivalent - i.e. that the following diagram commutes:
\begin{equation}\label{eq:pentdiagram}
\centering
\begin{tikzpicture}[scale=0.5]
	\begin{pgfonlayer}{nodelayer}
		\node [style=none] (0) at (-6.5, 3) {};
		\node [style=none] (1) at (-5.5, 3) {};
		\node [style=none] (2) at (-4.5, 3) {};
		\node [style=none] (3) at (-3.5, 3) {};
		\node [style=none] (4) at (-6, 2.5) {};
		\node [style=none] (5) at (-5.5, 2) {};
		\node [style=none] (6) at (-5, 1.5) {};
		\node [style=none] (7) at (-5, 0.5) {};
		\node [style=none] (8) at (-5, 2.5) {};
		\node [style=none] (9) at (-4, 2.5) {};
		\node [style=none] (10) at (-4.5, 2) {};
		\node [style=none] (11) at (-4, -1) {};
		\node [style=none] (12) at (-3, -1) {};
		\node [style=none] (13) at (-2, -1) {};
		\node [style=none] (14) at (-1, -1) {};
		\node [style=none] (15) at (-3.5, -1.5) {};
		\node [style=none] (16) at (-3, -2) {};
		\node [style=none] (17) at (-2.5, -2.5) {};
		\node [style=none] (18) at (-2.5, -3.5) {};
		\node [style=none] (19) at (-2.5, -1.5) {};
		\node [style=none] (20) at (-1.5, -1.5) {};
		\node [style=none] (21) at (-2, -2) {};
		\node [style=none] (22) at (-1.5, 6) {};
		\node [style=none] (23) at (-0.5, 6) {};
		\node [style=none] (24) at (0.5, 6) {};
		\node [style=none] (25) at (1.5, 6) {};
		\node [style=none] (26) at (-1, 5.5) {};
		\node [style=none] (27) at (-0.5, 5) {};
		\node [style=none] (28) at (0, 4.5) {};
		\node [style=none] (29) at (0, 3.5) {};
		\node [style=none] (30) at (0, 5.5) {};
		\node [style=none] (31) at (1, 5.5) {};
		\node [style=none] (32) at (0.5, 5) {};
		\node [style=none] (33) at (1, -1) {};
		\node [style=none] (34) at (2, -1) {};
		\node [style=none] (35) at (3, -1) {};
		\node [style=none] (36) at (4, -1) {};
		\node [style=none] (37) at (1.5, -1.5) {};
		\node [style=none] (38) at (2, -2) {};
		\node [style=none] (39) at (2.5, -2.5) {};
		\node [style=none] (40) at (2.5, -3.5) {};
		\node [style=none] (41) at (2.5, -1.5) {};
		\node [style=none] (42) at (3.5, -1.5) {};
		\node [style=none] (43) at (3, -2) {};
		\node [style=none] (44) at (3.5, 3) {};
		\node [style=none] (45) at (4.5, 3) {};
		\node [style=none] (46) at (5.5, 3) {};
		\node [style=none] (47) at (6.5, 3) {};
		\node [style=none] (48) at (4, 2.5) {};
		\node [style=none] (49) at (4.5, 2) {};
		\node [style=none] (50) at (5, 1.5) {};
		\node [style=none] (51) at (5, 0.5) {};
		\node [style=none] (52) at (5, 2.5) {};
		\node [style=none] (53) at (6, 2.5) {};
		\node [style=none] (54) at (5.5, 2) {};
		\node [style=none] (55) at (-5, 1.5) {};
		\node [style=none] (56) at (-4.5, 1) {};
		\node [style=none] (57) at (-3.25, -0.5) {};
		\node [style=none] (58) at (3.25, -0.5) {};
		\node [style=none] (59) at (4.5, 1) {};
		\node [style=none] (60) at (-3, 3.5) {};
		\node [style=none] (61) at (-1.25, 4.25) {};
		\node [style=none] (62) at (1.25, 4.25) {};
		\node [style=none] (63) at (3, 3.5) {};
		\node [style=none] (64) at (-1, -2.5) {};
		\node [style=none] (65) at (1, -2.5) {};
	\end{pgfonlayer}
	\begin{pgfonlayer}{edgelayer}
		\draw (0.center) to (55.center);
		\draw (55.center) to (7.center);
		\draw (55.center) to (3.center);
		\draw (5.center) to (2.center);
		\draw (4.center) to (1.center);
		\draw (22.center) to (28.center);
		\draw (28.center) to (29.center);
		\draw (24.center) to (31.center);
		\draw (25.center) to (28.center);
		\draw (26.center) to (23.center);
		\draw (44.center) to (50.center);
		\draw (50.center) to (51.center);
		\draw (50.center) to (47.center);
		\draw (45.center) to (54.center);
		\draw (46.center) to (53.center);
		\draw (11.center) to (17.center);
		\draw (17.center) to (18.center);
		\draw (17.center) to (14.center);
		\draw (12.center) to (19.center);
		\draw (13.center) to (16.center);
		\draw (33.center) to (39.center);
		\draw (39.center) to (40.center);
		\draw (39.center) to (36.center);
		\draw (34.center) to (43.center);
		\draw (41.center) to (35.center);
		\draw [->] (56.center) to (57.center);
		\draw [->] (60.center) to (61.center);
		\draw [->] (62.center) to (63.center);
		\draw [->] (58.center) to (59.center);
		\draw [->] (64.center) to (65.center);
	\end{pgfonlayer}
\end{tikzpicture}
\end{equation}\\
The commutation of this diagram defines a set of algebraic equations on the $F$-symbols:
\begin{equation}\label{eq:pentagon}
(F_d^{e,g,h})_{i,c}(F_d^{a,b,c})_{e,f}=\sum_{j\in\mathcal{C}}(F_i^{a,b,g})_{e,j}(F_d^{a,j,h})_{i,f}(F_f^{b,g,h})_{j,c}.
\end{equation}
Solving these equations \textit{categorifies} the fusion ring - that is, a fusion ring and self-consistent $F$-symbols define a fusion category $\mathcal{C}$. Solutions to the pentagon equations for a given fusion ring are non-unique. Introducing gauge variables, the pentagon equations \eqref{eq:pentagon} are invariant under the transformation 
\begin{equation}
(F_d^{a,b,c})_{e,f}\rightarrow\frac{g_{a,b,e}g_{e,c,d}}{g_{a,f,d}g_{b,c,f}}(F_d^{a,b,c})_{e,f}.
\end{equation}
Fusion categories defined by gauge equivalent $F$-symbols are considered equivalent. There may be several gauge-inequivalent solutions to the pentagon equations defined by a given fusion ring, and there may exist no solution. In this case the fusion ring is not categorifiable.
\par
In general, solving the $F$-symbols for a given fusion ring is very difficult. There are $n^6$ $F$-symbols (where $n$ is the rank of the fusion ring), and while many are automatically zero by the fusion rules, the number of nontrivial pentagon equations grows rapidly with $n$. For example, the $\mathfrak{psu}(2)_6$ fusion ring defines $\approx400,000$ pentagon equations. Solving such a large set of order 3 equations in $\leq\mathcal{O}(n^6)$ variables is a difficult task, and for this reason, rank $\geq8$ fusion rings have not yet been systematically categorified. However, for rank $\leq7$, categorifications of fusion rings have been computed and catalogued \cite{gert_thesis, vercleyen_anyonica_2025, vercleyen_slingerland_anyonwiki_2025}.
\par
If, in some gauge, the $F$-symbols of a fusion category have $(F_d^{a,b,c})^{\dagger}=(F_d^{a,b,c})^{-1}$, the fusion category is called unitary. Although requiring unitarity of $F$-symbols is often a natural choice, the results presented here do not require unitarity.
\par
The quantum dimension of an object $a$ in a fusion category is defined as
\begin{align}
    \text{dim}\hspace{0.05cm}a=\left|(F_{a}^{a,\bar{a},a})_{1,1}\right|^{-1}.
\end{align}
For unitary fusion categories, $\text{dim}\hspace{0.05cm}a=d_{FP,a}$, where $d_{FP,a}$ is the Frobenius-Perron dimension of the object, defined as the largest eigenvalue of the matrix defined by the constants in the fusion rules, $N_{ab}^c$, where the $b$ and $c$ indices are taken to be matrix indices. In this case, the sign of $(F_{a}^{a,\bar{a},a})_{1,1}$ is denoted $\kappa_a$, and called the `Frobenius-Schur' indicator of object $a$. In the unitary case, the quantum dimensions encode the probability of objects $a$ and $b$ producing object $c$ upon fusion. Specifically, if $N_{ab}^c\neq 0$, then $\text{P}(a\otimes b\rightarrow c)=\frac{\text{dim}\hspace{0.05cm}c}{\text{dim}\hspace{0.05cm}a\hspace{0.05cm}\text{dim}\hspace{0.05cm}b}$.

\subsection{Anyonic chains} In this paper we are studying \textit{anyonic spin chains}, which describe the interactions of some non-invertible object $a$ in a fusion category $\mathcal{C}$. These are diagrammatically represented by a fusion tree with the interacting particle $a$ on each external edge. The internal edges are the dynamical variables of the system. As it is conventional to represent one-dimensional spin chains by a horizontal line of sites, the flow of causality in diagrams of anyonic spin chains is from top-left to bottom-right, and states of the system are of the form
\begin{equation}
\begin{tikzpicture}[scale=0.7]
	\begin{pgfonlayer}{nodelayer}
		\node [style=none] (0) at (6, -7.25) {};
		\node [style=none] (1) at (7.5, -7.25) {};
		\node [style=none] (2) at (4.5, -9.25) {};
		\node [style=none] (3) at (6, -9.25) {};
		\node [style=none] (4) at (7.5, -9.25) {};
		\node [style=none] (5) at (6.75, -9.75) {$x_2$};
		\node [style=none] (6) at (8.25, -9.75) {$x_{3}$};
		\node [style=none] (7) at (5.25, -9.75) {$x_{1}$};
		\node [style=none] (8) at (9, -9.25) {};
		\node [style=none] (9) at (6, -6.85) {$a$};
		\node [style=none] (10) at (7.5, -6.85) {$a$};
		\node [style=none] (11) at (0.5, -8.25) {$\ket{x_{1},x_2,\cdots,x_{L-1},x_L}=$};
		\node [style=none] (12) at (9.75, -9.25) {$\cdots$};
		\node [style=none] (13) at (12, -7.25) {};
		\node [style=none] (14) at (13.5, -7.25) {};
		\node [style=none] (15) at (10.5, -9.25) {};
		\node [style=none] (16) at (12, -9.25) {};
		\node [style=none] (17) at (13.5, -9.25) {};
		\node [style=none] (18) at (12.75, -9.75) {$x_{L-1}$};
		\node [style=none] (19) at (14.25, -9.75) {$x_{L}$};
		\node [style=none] (20) at (11.25, -9.75) {$x_{L-2}$};
		\node [style=none] (21) at (15, -9.25) {};
		\node [style=none] (22) at (12, -6.85) {$a$};
		\node [style=none] (23) at (13.5, -6.85) {$a$};
		\node [style=none] (24) at (15.75, -8.25) {,};
	\end{pgfonlayer}
	\begin{pgfonlayer}{edgelayer}
		\draw (4.center) to (1.center);
		\draw (3.center) to (0.center);
		\draw (2.center) to (8.center);
		\draw (17.center) to (14.center);
		\draw (16.center) to (13.center);
		\draw (15.center) to (21.center);
	\end{pgfonlayer}
\end{tikzpicture} \label{eq:chainbasis}
\end{equation}
where periodic boundary conditions are imposed. The fusion rules of the chosen fusion category $\mathcal{C}$ impose constraints on the objects allowed at neighbouring sites of the spin chain. In particular, we must have $x_{j+1} \in x_j \otimes a$ for each $j = 1,2,\dots,L$.\footnote{By $x_{j+1} \in x_j \otimes a$ we mean that $N^{x_{j+1}}_{x_j a}\neq 0$.} Hence, the dimension of the constrained Hilbert space does not grow as $(\text{rank } \mathcal{C})^L$, but as $(\text{dim}\hspace{0.05cm}a)^L$. For example, the golden chain Hilbert space grows as $\phi^L$, because the Fibonacci anyon's quantum dimension is the golden ratio $\phi = \frac{1+\sqrt{5}}{2}$.
\par
The Hamiltonians of our models will be defined by a sum of local projectors onto some object $b$. The local projector's action on an anyonic spin chain is defined by
\begin{equation}\label{eq:projectorform}
\begin{tikzpicture}[scale=0.7]
	\begin{pgfonlayer}{nodelayer}
		\node [style=none] (0) at (0, 0.15) {$=$};
		\node [style=none] (12) at (-0.875, 1.25) {};
		\node [style=none] (13) at (-2.375, 1.25) {};
		\node [style=none] (16) at (-1.125, 1) {};
		\node [style=none] (17) at (-2.125, 1) {};
		\node [style=none] (18) at (-1.625, 0.5) {};
		\node [style=none] (27) at (-2.375, 1.625) {$a$};
		\node [style=none] (28) at (-0.875, 1.625) {$a$};
		\node [style=none] (33) at (4.05, 0.1) {.};
		\node [style=none] (34) at (-1.625, -0.5) {};
		\node [style=none] (35) at (-0.5, -0.5) {};
		\node [style=none] (36) at (-2.75, -0.5) {};
		\node [style=none] (37) at (-2.175, -0.875) {$x_{i-1}$};
		\node [style=none] (38) at (-1.05, -0.875) {$x_{i+1}$};
		\node [style=none] (39) at (-1.25, 0.15) {$\tilde{x}_i$};
		\node [style=none] (40) at (3.3, 1.25) {};
		\node [style=none] (41) at (1.8, 1.25) {};
		\node [style=none] (42) at (3.05, 1) {};
		\node [style=none] (43) at (2.05, 1) {};
		\node [style=none] (44) at (2.55, 0.5) {};
		\node [style=none] (45) at (1.8, 1.625) {$a$};
		\node [style=none] (46) at (3.3, 1.625) {$a$};
		\node [style=none] (47) at (2.55, -0.5) {};
		\node [style=none] (48) at (3.675, -0.5) {};
		\node [style=none] (49) at (1.425, -0.5) {};
		\node [style=none] (50) at (2, -0.875) {$x_{i-1}$};
		\node [style=none] (51) at (3.125, -0.875) {$x_{i+1}$};
		\node [style=none] (52) at (2.925, 0.15) {$\tilde{x}_i$};
		\node [style=none] (53) at (0.875, 0.15) {$\delta_{\tilde{x}_i,b}$};
		\node [style=none] (54) at (-3.25, 0.15) {$p_i^{(a,b)}$};
	\end{pgfonlayer}
	\begin{pgfonlayer}{edgelayer}
		\draw (16.center) to (18.center);
		\draw (18.center) to (17.center);
		\draw (17.center) to (13.center);
		\draw (16.center) to (12.center);
		\draw (18.center) to (34.center);
		\draw (36.center) to (34.center);
		\draw (34.center) to (35.center);
		\draw (42.center) to (44.center);
		\draw (44.center) to (43.center);
		\draw (43.center) to (41.center);
		\draw (42.center) to (40.center);
		\draw (44.center) to (47.center);
		\draw (49.center) to (47.center);
		\draw (47.center) to (48.center);
	\end{pgfonlayer}
\end{tikzpicture}
\end{equation}
Therefore, acting on the canonical basis where the dynamic variables are horizontal edges and the $a$ particles are vertical edges between them, the action of the projector is defined by locally changing basis, projecting, and reverting the change of basis:
\begin{equation}
\begin{tikzpicture}[scale=0.7]
	\begin{pgfonlayer}{nodelayer}
		\node [style=none] (70) at (-1.75, -2) {};
		\node [style=none] (71) at (-3.25, -2) {};
		\node [style=none] (72) at (-1.75, -3.75) {};
		\node [style=none] (73) at (-3.25, -3.75) {};
		\node [style=none] (75) at (-3.25, -1.625) {$a$};
		\node [style=none] (76) at (-1.75, -1.625) {$a$};
		\node [style=none] (77) at (-2.5, -3.75) {};
		\node [style=none] (78) at (-0.5, -3.75) {};
		\node [style=none] (79) at (-4.5, -3.75) {};
		\node [style=none] (80) at (-3.875, -4.125) {$x_{i-1}$};
		\node [style=none] (81) at (-1.125, -4.125) {$x_{i+1}$};
		\node [style=none] (82) at (-2.5, -4.125) {$x_i$};
		\node [style=none] (83) at (-5, -3.1) {$p_i^{(a,b)}$};
		\node [style=none] (84) at (0, -3.1) {$=$};
		\node [style=none] (85) at (10.65, -3.1) {$=(F_{x_{i+1}}^{x_{i-1},a,a})_{x_i,b}$};
		\node [style=none] (86) at (8.025, -2) {};
		\node [style=none] (87) at (6.525, -2) {};
		\node [style=none] (88) at (7.775, -2.25) {};
		\node [style=none] (89) at (6.775, -2.25) {};
		\node [style=none] (90) at (7.275, -2.75) {};
		\node [style=none] (91) at (6.525, -1.625) {$a$};
		\node [style=none] (92) at (8.025, -1.625) {$a$};
		\node [style=none] (94) at (7.275, -3.75) {};
		\node [style=none] (95) at (8.4, -3.75) {};
		\node [style=none] (96) at (6.15, -3.75) {};
		\node [style=none] (97) at (6.725, -4.125) {$x_{i-1}$};
		\node [style=none] (98) at (7.85, -4.125) {$x_{i+1}$};
		\node [style=none] (99) at (7.65, -3.1) {$\tilde{x}_i$};
		\node [style=none] (100) at (14.7, -2) {};
		\node [style=none] (101) at (13.2, -2) {};
		\node [style=none] (102) at (14.45, -2.25) {};
		\node [style=none] (103) at (13.45, -2.25) {};
		\node [style=none] (104) at (13.95, -2.75) {};
		\node [style=none] (105) at (13.2, -1.625) {$a$};
		\node [style=none] (106) at (14.7, -1.625) {$a$};
		\node [style=none] (107) at (13.95, -3.75) {};
		\node [style=none] (108) at (15.075, -3.75) {};
		\node [style=none] (109) at (12.825, -3.75) {};
		\node [style=none] (110) at (13.4, -4.125) {$x_{i-1}$};
		\node [style=none] (111) at (14.525, -4.125) {$x_{i+1}$};
		\node [style=none] (112) at (14.325, -3.05) {$b$};
		\node [style=none, align=left] (151) at (3.125, -3.2) {$\sum\limits_{\tilde{x}_i}(F_{x_{i+1}}^{x_{i-1},a,a})_{x_i,\tilde{x}_i}p_i^{(a,b)}$};
		\node [style=none] (154) at (0, -6.575) {$=$};
		\node [style=none, align=left] (170) at (4.025, -6.675) {$\sum\limits_{x_i'}(F_{x_{i+1}}^{x_{i-1},a,a})_{x_i,b}(F_{x_{i+1}}^{x_{i-1},a,a})_{b,x_i'}^{-1}$};
		\node [style=none] (171) at (10.65, -5.475) {};
		\node [style=none] (172) at (9.15, -5.475) {};
		\node [style=none] (173) at (10.65, -7.225) {};
		\node [style=none] (174) at (9.15, -7.225) {};
		\node [style=none] (175) at (9.15, -5.1) {$a$};
		\node [style=none] (176) at (10.65, -5.1) {$a$};
		\node [style=none] (177) at (9.9, -7.225) {};
		\node [style=none] (178) at (11.9, -7.225) {};
		\node [style=none] (179) at (7.9, -7.225) {};
		\node [style=none] (180) at (8.525, -7.6) {$x_{i-1}$};
		\node [style=none] (181) at (11.275, -7.6) {$x_{i+1}$};
		\node [style=none] (182) at (9.9, -7.6) {$x_i'$};
		\node [style=none] (183) at (12.15, -6.625) {.};
	\end{pgfonlayer}
	\begin{pgfonlayer}{edgelayer}
		\draw (73.center) to (71.center);
		\draw (72.center) to (70.center);
		\draw (79.center) to (77.center);
		\draw (77.center) to (78.center);
		\draw (88.center) to (90.center);
		\draw (90.center) to (89.center);
		\draw (89.center) to (87.center);
		\draw (88.center) to (86.center);
		\draw (90.center) to (94.center);
		\draw (96.center) to (94.center);
		\draw (94.center) to (95.center);
		\draw (102.center) to (104.center);
		\draw (104.center) to (103.center);
		\draw (103.center) to (101.center);
		\draw (102.center) to (100.center);
		\draw (104.center) to (107.center);
		\draw (109.center) to (107.center);
		\draw (107.center) to (108.center);
		\draw (174.center) to (172.center);
		\draw (173.center) to (171.center);
		\draw (179.center) to (177.center);
		\draw (177.center) to (178.center);
	\end{pgfonlayer}
\end{tikzpicture} \label{eq:Paction}
\end{equation}
This can be written in more compact notation as
\begin{equation}\label{eq:projdef}
p_i^{(a,b)}\ket{x_{i-1},x_i,x_{i+1}}=\sum_{x_i'}(F_{x_{i+1}}^{x_{i-1},a,a})_{x_i,b}(F_{x_{i+1}}^{x_{i-1},a,a})_{b,x_i'}^{-1}\ket{x_{i-1},x_i',x_{i+1}}.
\end{equation}
Since these operators are projectors, they satisfy $(p_i^{(a,b)})^2 = p_i^{(a,b)}$. We also note that since one fusion outcome is guaranteed, the local projectors satisfy the relation
\begin{equation}\label{eq:projsum}
\sum_{b \in a\otimes a} p_i^{(a,b)} = 1.
\end{equation}
A general anyonic chain describing the fusion of an object $a$ in a fusion category $\mathcal{C}$ is built as a linear combination of such projectors, summed over the whole chain:
\begin{equation}\label{eq:linearcomb}
\mathrm{P}^{\mathcal{C}}_a = \sum_{b \in a\otimes a} \alpha_b \sum_{i = 1}^L p_i^{(a,b)},
\end{equation}
where $\alpha_b \in \mathbb{C}$. The results of the present paper apply to the identity projector, i.e.\ we take $\alpha_b \propto \delta_{b, 1}$. To summarise, the steps taken to define an anyonic chain are:
\begin{itemize}
    \item Choose a fusion category $\mathcal{C}$ from which to build the system, i.e. a set of fusion rules and a corresponding solution to the pentagon equations \eqref{eq:pentagon}, $(F_d^{a,b,c})_{e,f}$.
    
    \item Choose a non-invertible object $a$ in the fusion category, which takes the role of the particle being fused. This object will lie on each of the external edges of the fusion tree  \eqref{eq:chainbasis}. This defines a constrained Hilbert space with dimension $\sim (\text{dim}\hspace{0.05cm}a)^L$.
    
 \item For each $b \in a\otimes a$, form the projectors $p_{i}^{(a,b)}$ defined in \eqref{eq:projdef} and choose an appropriate linear combination to study \eqref{eq:linearcomb}.
    
\end{itemize}

\subsection{Examples}
\label{sec:examples}

\subsubsection{$\mathfrak{psu}(2)_5$}\label{sec:psu25}
As a simple example, we consider the fusion algebra $\mathfrak{psu}(2)_5$. In this fusion algebra there are three objects, which we denote by $1, X, Y$. They obey the fusion rules
\begin{align}\label{eq:fusionrulespsu}
&X \otimes X = 1 \oplus Y, \notag \\ &Y\otimes Y = 1 \oplus X \oplus Y,\notag \\ &X\otimes Y = Y\otimes X = X \oplus Y,
\end{align}
and $1 \otimes a = a \otimes 1 = a$ for all $a \in \mathfrak{psu}(2)_5$. For the set of fusion rules \eqref{eq:fusionrulespsu}, there are three independent solutions to the pentagon identity \eqref{eq:pentagon}. One of these is unitary, and taking this solution leads to the quantum dimensions 

\begin{equation}
\text{dim}\hspace{0.05cm}1 = 1, \qquad \text{dim}\hspace{0.05cm}X = \frac{\sin \frac{5 \pi}{7}}{\sin \frac{\pi}{7}}, \qquad \text{dim}\hspace{0.05cm}Y = \frac{\sin \frac{3 \pi}{7}}{\sin \frac{\pi}{7}}.
\end{equation}
We note that in this fusion category, $\text{dim}\hspace{0.05cm}X<2$ and $\text{dim}\hspace{0.05cm}Y>2$. The $F$-symbols are available for example in \cite{Chang:2018iay} or on AnyonWiki \cite{vercleyen_slingerland_anyonwiki_2025}. The $F$-symbols can often be expressed in some gauge in terms of the quantum dimensions of the contributing objects. For example, in the unitary gauge provided by AnyonWiki the $F$-symbol $F_Y^{Y,Y,Y}$ can be expressed as
\begin{equation}
F_Y^{Y,Y,Y} = \begin{pmatrix} \frac{1}{d_Y} & \frac{\sqrt{d_X}}{d_Y} & \frac{1}{\sqrt{d_Y}} \\   \frac{\sqrt{d_X}}{d_Y}  & \frac{d_X}{d_Y^2} & -\left(\frac{\sqrt{d_X}}{\sqrt{d_Y}}\right)^3  \\  \frac{1}{\sqrt{d_Y}} &  -\left(\frac{\sqrt{d_X}}{\sqrt{d_Y}}\right)^3 & \frac{1}{d_Y^2} \end{pmatrix},
\end{equation}
where we abbreviated $d_X \equiv  \text{dim}\hspace{0.05cm}X$ and $d_Y\equiv  \text{dim}\hspace{0.05cm}Y $. There are two non-trivial anyonic chains one can form from this fusion category; one by taking the external object $a = X$, and one by taking $a = Y$. Taking $a = X$ leads to a `spin-1/2' chain with two non-vanishing local projectors, $p^{(X, 1)}_i$ and $p^{(X, Y)}_i$, of which only one is independent due to \eqref{eq:projsum}. Taking $a = Y$ leads to a `spin-1' chain with three non-vanishing local projectors,  $p^{(Y, 1)}_i$,  $p^{(Y, X)}_i$, and $p^{(Y, Y)}_i$, of which two are independent. In this case the dimension of the Hilbert space scales as $(d_Y)^L \sim 2.25^L$. For $L = 4$ there are 26 allowed states out of the possible $3^4 = 81$.\footnote{For example, $\ket{YXY1}$ is an allowed state, but $\ket{XYY1}$ is not, due to periodicity and $1 \notin Y\otimes X$.} There are several interesting combinations of projectors to study in this case. For example, there are three integrable spin chains which are realised as linear combinations $p^{(Y, X)}_i+\alpha p^{(Y, Y)}_i$ \cite{Vernier_2017}. The case $\alpha = 1$ gives the Temperley-Lieb chain, while the two cases $\alpha \neq 1$ realise integrable models based on the Birman-Murakami-Wenzl algebra.

\subsubsection{Haagerup fusion category}\label{sec:haagerup}
Another example is the Haagerup fusion category \cite{haagerup}, which has six objects $1, b, b^2, \rho, b\rho, b^2\rho$. The objects $1, b, b^2$ are invertible and $\rho,b\rho, b^2\rho$ are non-invertible. The non-trivial fusion rules are
\begin{equation}\label{eq:fusionhag}
\rho \otimes \rho = 1\oplus \rho \oplus b\rho \oplus b^2\rho, \qquad \rho\otimes b = b^2\otimes \rho,
\end{equation}
and $b^3 = 1$.\footnote{For example, $b\otimes b = b^2, b\otimes b^2 = b^2 \otimes b = 1$, and $b\otimes \rho = b\rho$.} In this case there are four independent solutions to the pentagon equations \eqref{eq:pentagon}. Two of these are unitary, and are typically denoted as $\mathcal{H}_2$ and $\mathcal{H}_3$. The $F$-symbols for these categories were presented in \cite{Huang:2020lox}. For definiteness we consider $\mathcal{H}_3$ in this paper, although the identity projectors corresponding to $\mathcal{H}_2$ and $\mathcal{H}_3$ are isomorphic.\footnote{The other projectors, for example $p_i^{(\rho,\rho)}$, differ between $\mathcal{H}_2$ and $\mathcal{H}_3$.} In $\mathcal{H}_3$ the quantum dimensions of the objects are 
\begin{equation}
\text{dim}\hspace{0.05cm}1 = \text{dim}\hspace{0.05cm}b = \text{dim}\hspace{0.05cm}b^2 = 1, \qquad \text{dim}\hspace{0.05cm}\rho = \text{dim}\hspace{0.05cm} b\rho= \text{dim}\hspace{0.05cm}b^2\rho = \frac{3+\sqrt{13}}{2} \sim 3.3.
\end{equation}
We construct anyonic chains in this fusion category by taking $a = \rho$. This defines a Hilbert space of dimension $\sim 3.3^L$, and four projectors $p_i^{(\rho, 1)}, p_i^{(\rho,\rho)}, p_i^{(\rho,b\rho)}, p_i^{(\rho, b^2\rho)}$, of which three are independent due to \eqref{eq:projsum}. The operator $p_i^{(\rho,\rho)}$ has been shown to be critical \cite{Huang:2021nvb}, and integrable generalisations found in \cite{Corcoran:2024eeh}.

\subsubsection{Fib$\times$Fib}\label{sec:FibFib}
As a final example, we consider the category constructed as a product of two individual Fibonacci categories. The usual Fibonacci category consists of two objects $1$ and $\tau$, with a non-trivial fusion rule $\tau\otimes \tau = 1\oplus \tau$ \cite{Trebst_2008}. There are two solutions to the pentagon equations \eqref{eq:pentagon}, one of which is unitary and gives the golden chain \cite{Feiguin:2006ydp}, one of which is non-unitary and gives the Lee-Yang chain \cite{Ardonne_2011}.

The product fusion category Fib$\times$ Fib consists of four objects $\boldsymbol{1}=(1,1),\boldsymbol{2}=(1,\tau),\boldsymbol{3}= (\tau,1),\boldsymbol{4}= (\tau,\tau)$. The fusion rules can be derived from those of the Fibonacci fusion category:
\begin{align}
 &\boldsymbol{2}\otimes \boldsymbol{2} = \boldsymbol{1} \oplus \boldsymbol{2},  &\boldsymbol{3}\otimes \boldsymbol{3} = \boldsymbol{1} \oplus \boldsymbol{3},\notag \\
& \boldsymbol{2}\otimes \boldsymbol{3} =   \boldsymbol{3}\otimes \boldsymbol{2} = \boldsymbol{4}, &\boldsymbol{2}\otimes \boldsymbol{4} =\boldsymbol{4}\otimes \boldsymbol{2} = \boldsymbol{3}\oplus\boldsymbol{4},\notag\\
&  \boldsymbol{3}\otimes \boldsymbol{4}=\boldsymbol{4}\otimes \boldsymbol{3}  = \boldsymbol{2}\oplus\boldsymbol{4}, &\boldsymbol{4}\otimes \boldsymbol{4} =\boldsymbol{1}\oplus\boldsymbol{2}\oplus\boldsymbol{3}\oplus\boldsymbol{4}.
\end{align}
$F$-symbols for Fib$\times$Fib can be obtained as products of $F$-symbols of Fib (see Appendix \ref{sec:appC}). Taking the unitary solution, the only choice of external object which gives models independent from the golden chain is $a = \boldsymbol{4}$. In this case there are four projectors $p_i^{\boldsymbol{4},\boldsymbol{j}}$ for $j=\boldsymbol{1},\dots,\boldsymbol{4}$, of which 3 are independent due to \eqref{eq:projsum}. The quantum dimension of the objects are
\begin{equation}
\text{dim}\hspace{0.05cm}\boldsymbol{1} = 1, \qquad \text{dim}\hspace{0.05cm}\boldsymbol{2} = \text{dim}\hspace{0.05cm}\boldsymbol{3} =\phi \qquad \text{dim}\hspace{0.05cm}\boldsymbol{4} = \phi + 1,
\end{equation}
where $\phi = \frac{1+\sqrt{5}}{2}$. Therefore the projectors $p_i^{\boldsymbol{4},\boldsymbol{j}}$ act on a Hilbert space of dimension $\sim (\phi+1)^L$. Anyonic chains based on products of Fibonacci categories were recently studied in \cite{Antunes:2025huk}. One such chain, claimed to be critical in \cite{Antunes:2025huk}, is shown in Appendix \ref{sec:appC} to coincide with the choice  $p_i^{\boldsymbol{4},\boldsymbol{1}}$ above. Chains based on coupled Temperley-Lieb models were also considered in \cite{Fendley_2008,Vernier_2014} The underlying physics will be discussed in section \ref{sec:gap}, in light of the underlying integrability.

\section{Temperley-Lieb and Integrable Models}\label{sec:TL}
In this section we recall the defining relations of the Temperley-Lieb algebra \cite{TemperleyLieb1971} and review how it can be used to construct integrable Hamiltonians. We show that anyonic chains which project to the identity fusion channel are always Temperley-Lieb, and thus define integrable spin chains. We describe the relation between the fusion category identity projectors and the ADE models considered previously by Pasquier. 
\subsection{Temperley-Lieb Structure of Models.}\label{par:TLStructure}

The Temperley-Lieb algebra is defined by a set of local generators $X_i$ and a parameter $\delta$ which satisfy the following relations:
\begin{align}\label{eq:TL}
    &X_i^2=\delta X_i, \notag\\ &X_iX_{i\pm1}X_i=X_i, \notag\\ &X_iX_j=X_jX_i\text{ where }|i-j|\geq2.
\end{align}

In this section we sketch that the identity projectors in a fusion category, appropriately normalised, provide a representation of \eqref{eq:TL}. The full details of the proof are given in Appendix \ref{sec:appA}. In particular, given a fusion category $\mathcal{C}$ and a non-invertible object $a$ which is self-dual ($a\otimes a = 1\oplus\ldots$), we define the operator $X_i^{(a)}$ via 
 \begin{equation}\label{eq:Xdef}
 X_i^{(a)}=\frac{1}{(F_a^{a,a,a})_{1,1}^{-1}}p_i^{(a,1)}.
\end{equation}
The models we study numerically are these densities summed on a chain of length $L$:
\begin{equation}\label{eq:Hamiltonian}
H^a = -\sum_{i=1}^L X_i^{(a)}, 
\end{equation}
with periodic boundary conditions. For a given fusion category $\mathcal{C}$, there are as many non-trivial operators $X_i^{(a)}$ as there non-invertible self-dual objects $a$. For $\mathcal{C} = \text{Fib}$ there is one operator $X_i^{(a)}$ which is the golden chain density \cite{Feiguin:2006ydp}. The full theorem we prove in Appendix \ref{sec:appA} is

\medskip

\textbf{Theorem.} \textit{Given a fusion category $\mathcal{C}$ containing a non-invertible self-dual object $a$, the operators $X_i^{(a)}$ satisfy the defining relations \eqref{eq:TL} of the Temperley-Lieb algebra.
The Temperley-Lieb parameter is $\delta=\frac{1}{(F_a^{a,a,a})_{1,1}^{-1}}$, which is equal to $\kappa_ad_{FP,a}$ in the special case of unitary fusion categories.
}
\medskip

The first property of \eqref{eq:TL} is clear: since $p_i^{(a,1)}$ is a projector, we have
\begin{equation}
 (X_i^{(a)})^2 = \left(\frac{1}{(F_a^{a,a,a})_{1,1}^{-1}}p_i^{(a,1)}\right)^2 = \frac{1}{(F_a^{a,a,a})_{1,1}^{-2}}p_i^{(a,1)} = \frac{1}{(F_a^{a,a,a})_{1,1}^{-1}}X_i^{(a)}.
\end{equation}
The third property of \eqref{eq:TL} is also fairly straightforward. It is clear from \eqref{eq:Paction} that $p_i^{(a,1)}$ is a range-3 operator which acts diagonally on sites $i\pm 1$, so that
\begin{equation}
X_i^{(a)} = D_{i-1}\mathcal{O}_i\tilde{D}_{i+1},
\end{equation}
where $D, \tilde{D}$ are diagonal operators and $\mathcal{O}$ is generic. Then for $j = i+2$ we have\footnote{Since $X^{(a)}_i$ is range-3, $j=i+2$ is the only non-trivial case to check.}
\begin{equation}
X_i^{(a)}X_{i+2}^{(a)} - X_{i+2}^{(a)} X_i^{(a)} = D_{i-1}\mathcal{O}_i \tilde{D}_{i+1}D_{i+1}\mathcal{O}_{i+2} \tilde{D}_{i+3}- D_{i+1}\mathcal{O}_{i+2} \tilde{D}_{i+3}D_{i-1}\mathcal{O}_i \tilde{D}_{i+1} = 0,
\end{equation}
since $[D_{i+1},\tilde{D}_{i+1}]=0$. The second property of \eqref{eq:TL} less obvious, and in this case the normalisation $\frac{1}{(F_a^{a,a,a})_{1,1}^{-1}}$ of $X_{i}^{(a)}$ is crucial. If $\tilde{x}_i=1$ in \eqref{eq:projectorform}, the fusion rules imply that $x_{i-1}=x_{i+1}$. Therefore, $p_i^{(a,1)}=\delta_{x_{i-1},x_{i+1}}p_i^{(a,1)}$. This and the pentagon equation 
\begin{equation}
(F_{x_{i}}^{x_{i},a,a})_{1,x_{i-1}}^{-1}(F_{x_{i-1}}^{x_{i-1},a,a})_{x_i,1}=(F_a^{a,a,a})_{1,1}^{-1}
\end{equation}
 establish the second property (see Appendix \ref{sec:appA} for the full details).

As discussed in the following section, this structure allows us to reconstruct our models from XXZ models with $\Delta=\frac{1}{2}\delta$. Since XXZ models are critical for $-1 < \Delta \leq 1$, we identify our models with $\delta>2$ as gapped models in the long-chain limit. 

\subsection{Relation to TLI and ADE models}\label{par:TLI/ADE}
Our models are related to the Temperley-Lieb interaction (TLI) models introduced in \cite{owczarek1987class} and reviewed in \cite{Pearce:1990ila}. TLI models are a subset of interaction-round-a-face models which are defined by an adjacency graph, and have local Hamiltonians whose Temperley-Lieb parameter equal to the adjacency matrix's Frobenius-Perron eigenvalue. It was observed that such models are critical if $\delta<2$ and considering simply connected adjacency graphs without loops, only ADE-type adjacency graphs fit this criterion. This subset of TLI models were studied by Pasquier \cite{PASQUIER1987162,Pasquier:1987xj} and are known as critical ADE models.
\par
In Appendix \ref{sec:appB}, the details of how our models are related to TLI models and Pasquier's critical ADE models are laid out. In short, our models are equivalent to a critical ADE model when we choose a unitary categorification of our fusion ring with $\text{dim}\hspace{0.05cm}a=d_{FP,a}\implies\kappa_a=1$ and make a specific choice of gauge. When defining our models, gauge-inequivalent categorifications of a fusion ring can lead to models with different Temperley-Lieb parameters - such models do not correspond to a TLI model, but under a choice of gauge they can be viewed as being the model that would be defined by the TLI-style construction where an eigenvalue other than the Frobenius-Perron eigenvalue is chosen.
\par
As such, the result that our models are critical when $\delta<2$ is a generalisation of the observation that ADE models are a critical subset of TLI models. Specifically, in the language of TLI models, we generalise the result that ADE models are critical in three ways:
\begin{itemize}
    \item Adjacency graphs with loops may arise from the fusion rules on our anyonic chain. If the external object of our anyonic chain has $\text{dim}\hspace{0.05cm}a<2$ in the chosen categorification, we have critical models with this (non-ADE-type) adjacency graph. 
    \item We can define critical models with $\text{dim}\hspace{0.05cm}a<2$, and $\text{dim}\hspace{0.05cm}a\neq d_{FP,a}$. Such models are beyond the scope of the construction of TLI models, but in a certain gauge they can be viewed as a model which is produced by generalising the TLI model construction to allow for choice of eigenvector of the adjacency matrix.
    \item For a given adjacency graph, gauge-transformations of the $F$-symbols provide us with a continuum of models with the same Temperley-Lieb parameter. While the matrix elements of the Hamiltonian are not invariant under these gauge transformations, the energy spectrum is (this is discussed in Appendix \ref{sec:appA}).
\end{itemize}

\subsection{Integrability} The usual definition for quantum integrability is the existence of an infinite number of commuting charges $Q_i$, of which the Hamiltonian is one, acting on the Hilbert space $\mathcal{H}$.
Hamiltonians written in terms of Temperley-Lieb generators as \eqref{eq:Hamiltonian} have long been known to be integrable irrespectively of the representation used for the latter, and indeed Nienhuis and Huijgen \cite{nienhuis2021local} recently found closed form expressions for the conserved charges in terms of the TL generators. Written as such, their commutation can be proved in a purely algebraic manner using the TL definig  relations \eqref{eq:TL} only, and therefore holds in any representation.  We note in passing that  other families of integrable Hamiltonians can be constructed out of TL generators by allowing longer-range interactions, see eg. \cite{ikhlef2009temperley}. 

Given a representation of the TL algebra, a convenient way to establish integrability of the Hamiltonian \eqref{eq:Hamiltonian} and generate the conserved charges is through a family of mutually commuting transfer matrices $t(u):\mathcal{H}\rightarrow \mathcal{H}$, built from a Lax operator $\mathcal{L}(u)$ which obeys an appropriate Yang--Baxter equation. The transfer matrix associated with the original appearances of the TL algebra, in relation with the Potts \cite{fortuin1972random} or XXZ spin chain \cite{pasquier1990common}, is that of the six-vertex model \cite{baxter2016exactly}. For anyonic chains such as those considered in this paper, the transfer matrices are instead built from a restricted solild-on-solid (RSOS) model \cite{pasquier1988etiology}. An alternative way to construct such transfer matrices is the language of medium range integrability \cite{Gombor:2021nhn}, which we will now describe.
In the present case, the integrable models we consider are range-3 and act on constrained Hilbert spaces, which requires the usual range-2 Lax formulation of integrability to be tweaked slightly \cite{Corcoran:2024ofo}. In short, one can take the range-3 Lax operator

\begin{equation}
\check{\mathcal{L}}_i(u) = 1 + \frac{\sinh(u)}{\sinh(\eta-u)} X_i^{(a)},
\end{equation}
where $2 \cosh \eta = \delta$. Using the Temperley-Lieb algebra \eqref{eq:TL} satisfied by $X_i^{(a)}$, it is then straightforward to show that this Lax operator satisfies the Yang--Baxter equation \cite{Wang_1996}
\begin{equation}\label{eq:YBE}
\check{\mathcal{L}}_{i}(u-v)\check{\mathcal{L}}_{i+1}(u)\check{\mathcal{L}}_{i}(v) = \check{\mathcal{L}}_{i+1}(v)\check{\mathcal{L}}_{i}(u)\check{\mathcal{L}}_{i+1}(u-v).
\end{equation}
In order to define the higher charges, we first expose the full index set of the range-3 operator $\check{\mathcal{L}}_{i}\equiv \check{\mathcal{L}}_{i,i+1,i+2}$, and define the unchecked Lax operator
\begin{equation}
\mathcal{L}_{i,i+1,i+2}(u) = \mathcal{P}_{i,i+2}\mathcal{P}_{i+1,i+2}\check{\mathcal{L}}_{i,i+1,i+2}(u),
\end{equation}
where $\mathcal{P}_{i,j}$ is the permutation operator. The transfer matrix can be defined from this operator via
\begin{equation}\label{eq:transfer}
t(u) = \text{tr}_{a,b}[\mathcal{L}_{a,b,L}(u) \mathcal{L}_{a,b,L-1}(u) \cdots \mathcal{L}_{a,b,1}(u) ],
\end{equation}
where $a,b$ are auxiliary space indices. The full set of charges can then be obtained via logarithmic derivatives of \eqref{eq:transfer}:

\begin{equation}
Q_i = \frac{d^{i-1}}{du^{i-1}}\log t(u)\Big|_{u=0},
\end{equation}
where it is easy to prove that $Q_2 = \sum_{i=1}^L X_i^{(a)}$. The mutual commutation of this set of charges $[Q_i, Q_j] =0$ follows from the commutativity of transfer matrices
\begin{equation}
[t(u),t(v)] = 0,
\end{equation}
which itself follows from the $R\mathcal{L}\mathcal{L}$ relation
\begin{equation}
R_{A,B}(u,v)\mathcal{L}_{A,i}(u)\mathcal{L}_{B,i}(v) =  \mathcal{L}_{B,i}(v)\mathcal{L}_{A,i}(u) R_{A,B}(u,v),
\end{equation}
where we introduced doubled auxiliary space indices $A=(a_1,a_2), B=(b_1, b_2)$. Using the Yang--Baxter equation \eqref{eq:YBE} one can show that \cite{Gombor:2021nhn}
\begin{equation}
\check{R}_{12,34}(u,v) = \check{\mathcal{L} }_{123}(-v) \check{\mathcal{L} }_{234}(u-v)\check{\mathcal{L} }_{123}(u),
\end{equation}
is an appropriate $R$-matrix, where the unchecked $R$ can be recovered via $R_{A,B}(u,v)=\mathcal{P}_{A,B}\check{R}_{A,B}(u,v)$.

\section{Gapped Temperley-Lieb chains and finite size effects}\label{sec:gap}

We now specify to families of models which correspond to Temperley-Lieb parameters $\delta>2$ among the examples listed in Section \ref{sec:examples}. Specifically, we will consider the TL chains built from categories $\mathcal{C} = \mathcal{H}_3$, $\mathcal{C} = \text{Fib}\times \text{Fib}$, and $\mathcal{C} = \mathfrak{psu}(2)_k$ (where we take the spin-1 external object\footnote{We take $a$ to be the unique object in $\mathfrak{psu}(2)_k$ such that $a\otimes a$ contains three elements.}), for which the parameters $\delta$ read respectively 
\be
\begin{split}
\delta_{\mathcal{H}_3} &= \frac{3+\sqrt{13}}{2} \simeq 3.30278  \\
\delta_{\text{Fib}\times \text{Fib}} &=  \frac{3+\sqrt{5}}{2} \simeq 2.61803  \\ 
\delta_{\mathfrak{psu}(2)_k,\text{spin } 1} &= 1 + 2 \cos\left(\frac{2\pi}{k+2}\right) \,.
\end{split}
\ee 
For the $\mathfrak{psu}(2)_k$ case, $\delta>2$ for $k\geq 5$, and in the following we will focus on $k=5$ for concreteness.
Such chains have been considered in the recent literature  \cite{Corcoran:2024eeh, Antunes:2025huk, Gils_2013}, but in all cases the numerical analysis of the spectrum was  plagued by important finite-size corrections, obscuring the underlying gapped behaviour. In this section we shall therefore revisit the analysis of the low-energy spectrum by exploiting the underlying integrability of the models.

\subsection{Decomposition in terms of TL modules}
\label{sec:decompositionspectrum}

Since all models can be formulated in terms of Temperley-Lieb generators, their Hilbert space can be recast as a direct sum of irreducible TL representations. Having at hand such a decomposition is very helpful, as it fully determines the spectrum of the Hamiltonian \eqref{eq:Hamiltonian}.
Our first task is therefore to understand how for each model the Hilbert space decomposes into irreps.

The irreducible representations of the (periodic) TL algebra are usually expressed geometrically in terms of link diagrams \cite{graham1998representation,martin1994blob}. 
They also appear naturally in another well-known incarnation of the TL algebra, namely the (periodic) XXZ spin chain \cite{pasquier1990common}. Since the latter is naturally amenable to a Bethe ansatz treatment we will stick with it in the following, and refer to e.g.\ \cite{Fukai_2024} for a recent review of the correspondence with the graphical presentation.

The XXZ chain is defined on a set of $L$ spins-1/2, with total Hilbert space $\mathcal{H}_{\rm XXZ}= (\mathbb{C}^2)^{\otimes L}$. The TL generators act on pairs of consecutive spins as 
\begin{equation}
  X_j=h^{\rm XXZ}_{j,j+1}\,,
\end{equation}
where $h^{\rm XXZ}_{j,j+1}$ is a two-site operator given by \cite{pasquier1990common}
\begin{equation}
\label{eq:hTL}
h^{\rm XXZ}_{j,j+1}=-\frac{1}{2}\left[2e^{i\varphi/L}\sigma^+_j \sigma^-_{j+1}+2e^{-i\varphi/L}\sigma^-_j \sigma^+_{j+1}+\cosh\eta\big(\sigma^z_j\sigma^z_{j+1}-1\big)+ \sinh\eta\big(\sigma^z_j-\sigma^z_{j+1}\big)\right] \,, 
\end{equation}
being understood that we take periodic boundary conditions, namely $L+1 \equiv 1$. Here $\sigma^{x,y,z}_j$ are the usual Pauli matrices acting the $j$\textsuperscript{th} spin (and as identity elsewhere), and $\sigma^\pm = \frac{\sigma^x\pm i \sigma^y}{2}$.
The parameter $\eta$ in \eqref{eq:hTL} is related to the Temperley-Lieb parameter $\delta$ by
\be
\delta= 2 \Delta=2\cosh\eta \,,
\ee
which we have also related to the usual anisotropy parameter $\Delta$ of the XXZ chain. 
In the following we will sometimes use the notations $\eta_{\text{Fib}\times \text{Fib}}$, $\eta_{\mathcal{H}_3}$, $\eta_{\mathfrak{psu}(2)_k}$ (and similarly for $\Delta$) for the parameters attached to the various models under consideration.
The parameter $\varphi\in\mathbb{C}$, in turn, is the so-called twist parameter, and will be shown to play an important role in the following. In the representation \eqref{eq:hTL} the twist is distributed homogeneously over the entire chain; another possible equivalent representation, related to \eqref{eq:hTL} by a simple change of basis, is to have all dependence in the twist concentrated on the last density $h^{\rm XXZ}_{L,1}$, wherein $\varphi/L$ is replaced by $\varphi$ while it is set to zero in all other Hamiltonian densities.

The Temperley-Lieb operators \eqref{eq:hTL} commute with the global magnetization $S^z = \frac{1}{2} \sum_{j=1}^L \sigma^z$, and therefore the XXZ representation can be reduced by specifying the value of the latter. In fact, all the irreducible representations of the (periodic) TL algebra can be recovered by fixing the magnetization $m=S^z$ and the twist parameter (since the energies are invariant under a global flip of all spins, we may restrict to positive values of $m$). In the following, we will denote such representations (``modules'') by the doublet $(m,\varphi)$. The corresponding dimensions are given by binomial coefficients 
\be 
\dim (m,\varphi) =  \binom{L}{L/2-m}  \,.
\ee 
We will furthermore restrict ourselves to even values of $L$, in which case $m$ takes integer values. We note that we obtained these decompositions into TL modules numerically on a case-by-case basis. Proving these mappings rigorously or finding a general pattern which holds for all fusion categories is an interesting open question for the future.

\subsubsection{The Haagerup chain}
We first discuss the TL chain built from the Haagerup $\mathcal{H}_3$ fusion category, discussed in section \ref{sec:haagerup}. The dimension of the corresponding Hilbert space scales as $(\frac{3+\sqrt{13}}{2})^L$. We obtain the decomposition of the Hilbert space into irreducible modules by comparing the Hamiltonian spectrum for finite sizes with the XXZ spectrum at different twist angles in the various irreducible modules, and recomposing the former in terms of the latter. 
For $L=4,6,8$, we find respectively 
\begin{align} 
\mathcal{H}_{\mathcal{H}_3}^{(L=4)} =& (0,\varphi_0) \oplus 4 (0,\varphi_1) \oplus (0,\varphi_2) \oplus 3(1,0)\oplus 75(2,-) 
\\
\mathcal{H}_{\mathcal{H}_3}^{(L=6)} =& (0,\varphi_0) \oplus 4 (0,\varphi_1) \oplus (0,\varphi_2) \oplus 3(1,0)\oplus 39(2,0) \oplus 36(2,\pi) \oplus  687 (3,-)
\\
 \mathcal{H}_{\mathcal{H}_3}^{(L=8)} =& (0,\varphi_0) \oplus 4 (0,\varphi_1) \oplus (0,\varphi_2) \oplus 3(1,0)\oplus 39(2,0) \oplus 36(2,\pi) \oplus  231 (3,0) 
 \nonumber \\ 
 & \oplus  456 (3,2\pi/3) \oplus  5979  (4,-) \,,
\end{align} 
where the integer coefficients in front of some modules indicate the multiplicity with which they appear in the spectrum. 
The twists $\varphi_0$, $\varphi_1$, $\varphi_2$ are given by 
\be 
\cos\varphi_0 = \cosh (2\eta_{\mathcal{H}_3}) \,, \qquad \cos \varphi_1 = -\frac{1}{2} \,, \qquad \cos\varphi_2 = \frac{7-3\sqrt{13}}{4}  \,.
\label{twistsH}
\ee 
Note also that on each line the twist of the module with $m=L/2$ is left unspecified, as the latter corresponds to the trivial one-dimensional module where all TL generators are null, irrespectively of the twist.
An elementary check of the above decompositions is that the sum of dimensions of modules appearing  of the right-hand-side equals the total dimension of the Haagerup chain Hilbert space. For instance, for $L=4$, $(1+4+1)\binom{4}{2}+3\binom{4}{1}+(36+39)\binom{4}{0}=123$.

\begin{figure}
\begin{center}
\includegraphics[scale=0.8]{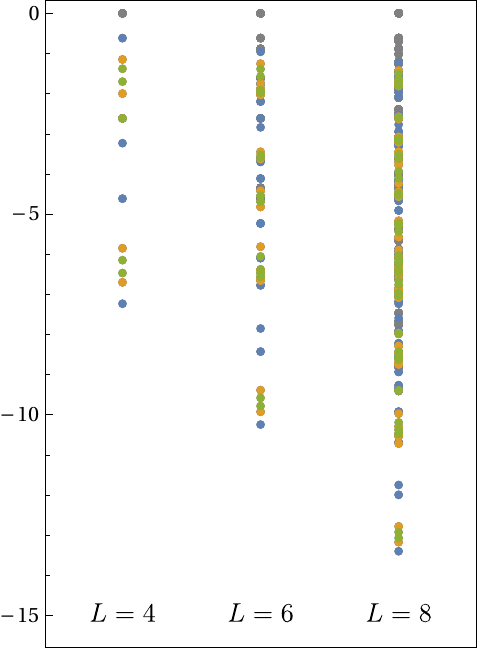}
\hspace{2cm}
\includegraphics[scale=0.8]{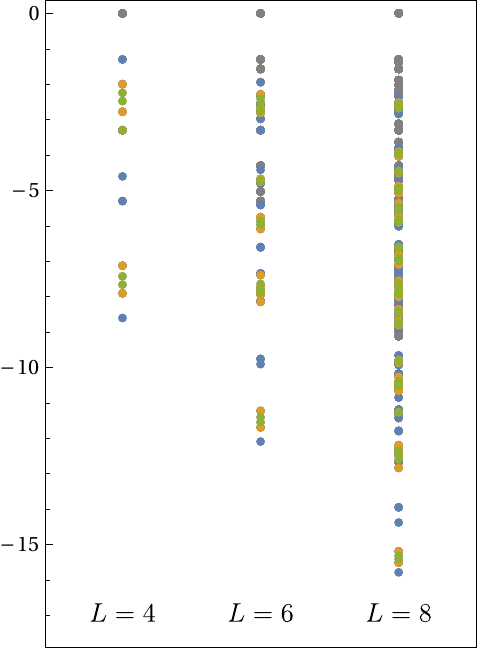}
\caption{Spectra of the Hamiltonian for the Fib$\times$Fib (left) and Haagerup (right) chains for $L=4,6,8$. We have indicated in blue (resp. yellow, green) the energy levels that can be recovered from the XXZ chain at zero magnetization and twists $\varphi_0$ (resp. $\varphi_1$, $\varphi_2$) as defined in the main text, see eqs \eqref{twistsFF} and \eqref{twistsH}. The remaining energy levels, marked in gray, are recovered by XXZ sectors with nonzero values of the magnetization.}  
\label{fig:spectra}
\end{center}
\end{figure}

\begin{figure}
\begin{center}
\includegraphics[scale=0.6]{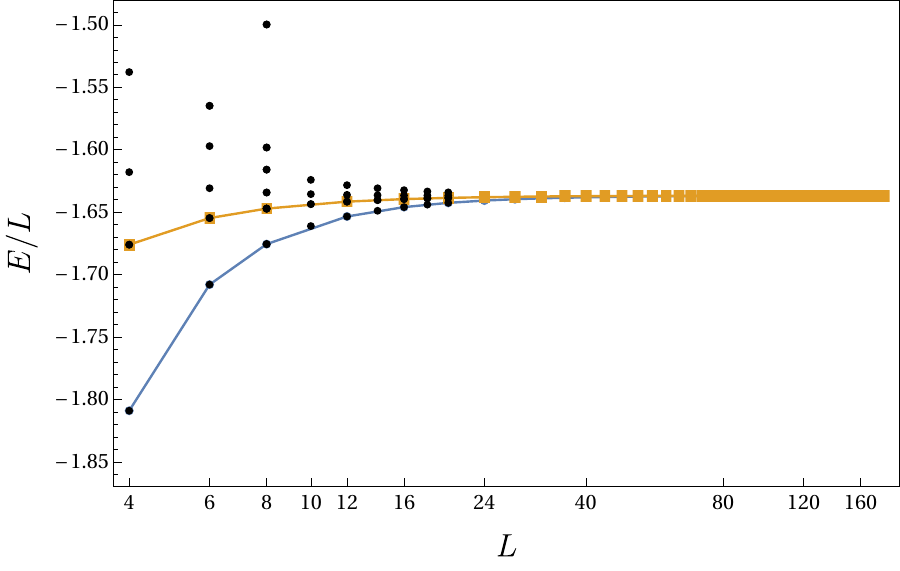}
\hspace{2cm}
\includegraphics[scale=0.6]{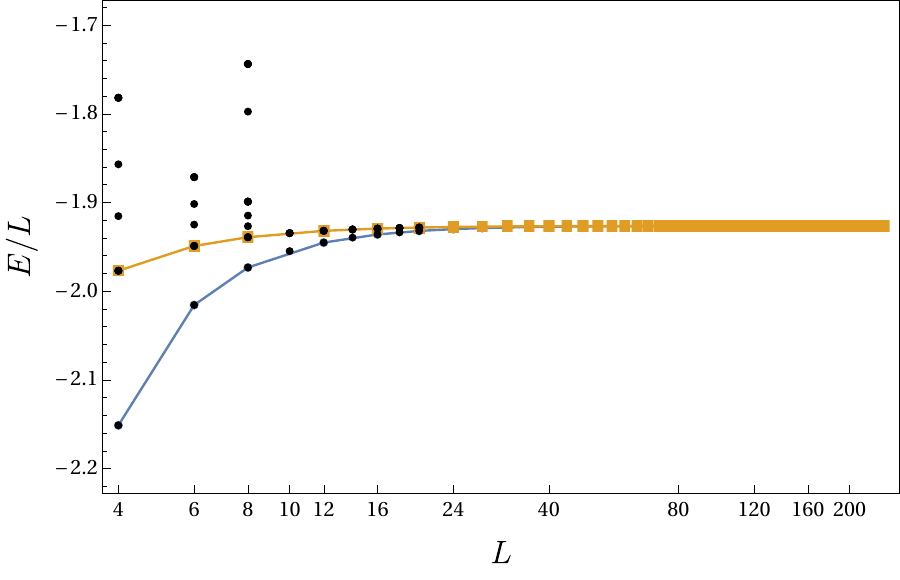}
\caption{Low-lying spectra of the Fib$\times$Fib (top) and Haagerup chains (bottom) for increasing system sizes: numerical data (obtained form exact diagonalization for $L$ up to $8$, and from DMRG for $L$ between $10$ and $20$) is shown as black dots, and the blue/orange dotted lines show the lowest energies $E_0(L)$ and $E_1(L)$ in the sectors $(0,\varphi_0)$ and $(0,\varphi_1)$, computed by Bethe ansatz.}  
\label{fig:spectralargeL}
\end{center}
\end{figure}

A clear pattern emerges from there: as $L$ is increased, modules with increasing values of $m$ (and twists which are rational multiples of $\pi$) are added to the decomposition. We therefore write for general even $L$
\be  \label{eq:haagerupdecomp}
\mathcal{H}_{\mathcal{H}_3} = (0,\varphi_0) \oplus 4 (0,\varphi_1) \oplus (0,\varphi_2) \oplus 3(1,0)\oplus 39(2,0) \oplus 36(2,\pi) \oplus  \ldots  \,,
\ee
where the dots indicate contributions from sectors with increasing magnetization, whose energy increases accordingly. 
In particular, the lowest energy levels are all contained in the $(0,\varphi_0)$, $(0,\varphi_1)$ and $(0,\varphi_2)$ sectors, see Figure \ref{fig:spectra} for data at $L=4,6,8$. 
Furthermore, we check for larger sizes using DMRG\footnote{We used ITensor \cite{Fishman_2022} to perform a check of the decomposition \eqref{eq:haagerupdecomp} and similar. We typically use a maximum bond dimension of 1500 and 30 sweeps up to length $L=20$. The Hilbert space constraints are enforced by adding an energy penalty for disallowed states.} that the ground state is obtained as the lowest energy level of the $(0,\varphi_0)$ sector, while the first excited state corresponds to the lowest energy level of the $(0,\varphi_1)$ sector, see Figure \ref{fig:spectralargeL}.

\subsubsection{The Fib$\times$Fib chain}

A similar analysis can be made for the Temperley-Lieb Fib$\times$Fib chain. We discussed this fusion category in section \ref{sec:FibFib} and in this case the dimension of the Hilbert space scales as $(\phi+1)^L$. We find that the Hilbert space can generically be decomposed as  
\be 
\mathcal{H}_{\text{Fib}\times \text{Fib}} = (0,\varphi_0) \oplus 2 (0,\varphi_1) \oplus (0,\varphi_2) \oplus (1,0) \oplus  11 (2, 0) 
 \oplus 10 (2,\pi) \oplus \dots 
\ee 
where the twists $\varphi_0$, $\varphi_1$, $\varphi_2$ are defined by 
\be 
\cos\varphi_0 = \cosh (2\eta_{\text{Fib}\times \text{Fib}}) \,, \qquad \cos \varphi_1 = -\frac{1}{2} \,, \qquad \cos\varphi_2 = \frac{3}{4}(1-\sqrt{5}) \,
\label{twistsFF}
\ee 
(note the $\eta$-dependence of $\varphi_0$, which is the same as in the Haagerup chain). 
Again, the lowest energies come from the zero magnetization sectors, see Figure \ref{fig:spectra} and \ref{fig:spectralargeL}. More precisely, as for the Haagerup chain, the energies of the ground state and first excited state are obtained as the lowest energies of the sectors $(0,\varphi_0)$ and $(0,\varphi_1)$ respectively.

\subsubsection{The $\mathfrak{psu}(2)_5$ chain}

We also consider the Temperley-Lieb chain constructed from the spin-$1$ object $Y$ in the $\mathfrak{psu}(2)_5$ fusion category (introduced in section \ref{sec:psu25}). In this case the Hilbert space scales as $2.25^L$. The Hilbert space of the $\mathfrak{psu}(2)_5$ chain can be decomposed in a similar way to the previous cases (more generally such a decomposition for arbitrary $\mathfrak{psu}(2)_k$ was given in \cite{Vernier_2017}).
We find 
\be 
\mathcal{H}_{\mathfrak{psu}(2)_5} = (0,\varphi_0) \oplus  (0,\varphi_1^\ast) \oplus (0,\varphi_2^\ast) \oplus 4 (2,0) \oplus  4 (2, \pi)   \oplus 7(3,0) \oplus 14 (3,2\pi/3)\oplus \ldots\,,
\ee 
where the twists $\varphi_0$, $\varphi_1^\ast$, $\varphi_2^\ast$ are now given by
\be 
\cos\frac{\varphi_0}{2} = \cosh (\eta_{\mathfrak{psu}(2)_5})
= \frac{1}{2} + \cos\left(\frac{2\pi}{7}\right)
 \,,~
\cos\frac{\varphi_1^\ast}{2} =  \frac{1}{2} + \cos\left(\frac{4\pi}{7}\right)
 \,,~
\cos\frac{\varphi_2^\ast}{2} =  \frac{1}{2} + \cos\left(\frac{6\pi}{7}\right) \,.
\label{twistsPSU}
\ee 
While the twist $\varphi_0$ has the same dependence 
in $\eta$ as for the Haagerup and Fib $\times$ Fib chains, namely $\cos\varphi_0=\cosh(2\eta)$, the twists $\varphi_1^\ast$ and $\varphi_2^\ast$ do not, hence the different notation.

\subsection{The low-energy spectrum}
\label{sec:lowenergy}

Interpreting the energies of the anyonic chains in terms of the XXZ chain enables us to use the Bethe ansatz solution of the latter in order to compute those energies for larger size. 
Before doing so, however, we can already reach a good understanding of the low-energy properties in light of the very well-known physics of the XXZ chain or, equivalently, of the closely related six-vertex model \cite{baxter2016exactly}. 

For real $\eta$ (equivalently, $\delta>2$, $\Delta>1$, which is the case for all three examples considered in this section), the XXZ chain is known to be gapped, and sits at the locus of a first order phase transition, which is that of the $Q$-states Potts model for $Q=\delta^2>4$ \cite{fortuin1972random}.
{
A qualitative understanding of the ground state and structure of low-energy excitations can be gained from perturbation theory at large coupling, $\Delta\to\infty$. To leading order the Hamiltonian is $H_{0}^{\rm XXZ}=\frac{\Delta}{2} \sum_j \sigma_j^z \sigma_{j+1}^{z}$, which has two degenerate ground states corresponding to the purely antiferromagnetic Néel and anti-Néel states. Excitations on top of the doubly degenerate ground state are built from domain walls between those two states, each domain wall coming at an energy cost $\Delta$. The next order contribution to the Hamiltonian  flips pairs of adjacent spins, and vanishes on the ground state manifold. The ground states therefore remain degenerate to first order in $1/\Delta$, and separated by a finite gap from the rest of the system. Similar observations hold at higher orders, and in order to lift the ground state degeneracy, one has to go to order $1/\Delta^L$. This results in the following properties of the spectrum: 
\begin{itemize} 
\item At a fixed value of the twist, the splitting between the two lowest energies is exponentially small in $L$, while higher energies are separated by a finite gap of order $\Delta$.
\item Changing the twist $\varphi$ by a finite amount results in an exponentially small shift in the ground state energy.
\end{itemize} 
}
The same conclusions hold beyond perturbation theory, for any $\Delta>1$. More quantitatively, one should expect that in the gapped phase the effect of a twist $\varphi$ (which, as discussed above, can be thought of as a local perturbation of magnitude $\varphi$ located on two sites of the chain) on the ground state energy decays as $e^{-L / \xi}$, where $\xi$ is the correlation length of the model, known from the exact solution \cite{baxter2016exactly} to be of the form 
\be 
\xi^{-1} = - \ln \left(2 e^{-\eta/2} \prod_{p=1}^\infty \left(\frac{1+ e^{-4 p \eta}}{1+ e^{-(4 p-2) \eta} } \right)^2 \right) \,. 
\label{correlationlength}
\ee 
For the models studied here the correlation lengths are 
\be 
\xi_{\text{Fib}\times \text{Fib}} \simeq 155.392 \,, \qquad \xi_{\mathcal{H}_3} \simeq 23.4085 \,, \qquad
\xi_{\mathfrak{psu}(2)_5} \simeq 5675.58 \,,
 \ee 
in other words they can be quite large, and lead to important finite size effects when studying systems of, say, a few dozen sites.  
In order to quantitatively observe the closure of the gap, we therefore use the underlying integrability of the models to compute the latter with Bethe ansatz.

\subsection{Bethe ansatz study of the gap}

As we saw in Sec. \ref{sec:decompositionspectrum}, the lowest energies of the Fib$\times$Fib, Haagerup and $\mathfrak{psu}(2)_5$ match with those of the periodic XXZ chain in the sector of zero magnetization, at twists $\varphi_0$ and $\varphi_1$ ($\varphi_1^\ast$ for $\mathfrak{psu}(2)_5$) defined respectively in eqs. \eqref{twistsFF}, \eqref{twistsH}, \eqref{twistsPSU}. 
Because of the similar expressions of the twists $\varphi_0$ and $\varphi_1$ for the Fib$\times$Fib and Haagerup chains, we shall here limit our study to these two cases. The case of $\mathfrak{psu}(2)_5$ (and, as a matter of fact, other chains associated with $\mathfrak{psu}(2)_k$ for $k\geq 5$) can be treated analogously, as shown in \cite{Vernier_2017}.
In the following, we will call the two corresponding energies $E_0(L)$ and $E_1(L)$ respectively. In fact, it will be useful to go beyond the Fib$\times$Fib and Haagerup points and consider for generic real $\eta$ the energies $E_0(L)$ and $E_1(L)$ defined as the ground states in the sectors $(0,\varphi_0)$, $(0,\varphi_1)$, where $\cos\varphi_0=\cosh(2 \eta)$ and $\cos(\varphi_1)=-\frac{1}{2}$.

In the Bethe ansatz, the eigenstates are constructed as interacting spin waves created on top of a reference state. They are characterised by a set of rapidities $\{\lambda_1,\dots,\lambda_N\}$, where $N$ is related to the magnetization $m=S^z$ through $N=L/2-m$, while the rapidities satisfy the so-called Bethe equations \cite{baxter2016exactly,takahashi1999thermodynamics}
\begin{equation}
e^{i\varphi}  
\left(  
  \frac{\sin(\lambda_j+i\eta/2)}{\sin(\lambda_j-i\eta/2)}
  \right)^L
  \prod_{k\ne j}   
  \frac{\sin(\lambda_j-\lambda_k-i\eta)}{\sin(\lambda_j-\lambda_k+i\eta)}=1.
  \label{eq:BAE}
\end{equation}

For a given such eigenstate, the corresponding eigenvalue of $H$ can be expressed as a sum over all rapidities,
\begin{equation}
  E=\sum_{j=1}^N \epsilon(\lambda_j) \,, 
  \label{energies1}
\end{equation}
where the single-rapidity energy is given by 
  \begin{equation}
  \epsilon(\lambda)=\frac{\sinh^2\eta}{\sin(\lambda+i\eta/2)\sin(\lambda-i\eta/2)}\,.
  \label{eq:logBAE}
\end{equation}
  
The structure of the ground state and low-lying excitations of the XXZ chain at zero (as well as for real) twist is well-known \cite{baxter2016exactly,takahashi1999thermodynamics,des1966anisotropic}: in the zero-magnetization sector, the ground state is associated for real $\eta$ with a set of $N=L/2$ real rapidities distributed along the interval $[-\pi/2,\pi/2]$. 
These may be characterized by a set of $N$ distinct integer quantum numbers $I_j$, whose definition follows from taking the logarithm form of eqs. \eqref{eq:BAE}, 
\be
L \phi_{\tfrac{\eta}{2}}(\lambda_j) - 
  \sum_{k\ne j}   \phi_{\eta}(\lambda_j-\lambda_k)
 =2 \pi I_j + \varphi \,, 
  \qquad \phi_\chi(\lambda) \equiv i \log\frac{\sin(\lambda+i \chi)}{\sin(\lambda-i \chi)} \,.
 \ee
For the ground state the integers $\{I_j\}$ are symmetrically packed around the origin, resulting for zero twist in a symmetric distribution of real Bethe roots,
while turning on a non-zero real twist results in an asymmetric distribution. Turning on an imaginary twist, this distribution deforms continuously and the rapidities becomes complex.  
{This is illustrated on Figure \ref{fig:Betheroots}, where we show the root distributions associated with the ground state at $\varphi=0$ and $\varphi=i$ (blue and green dots respectively) for a given system size and anisotropy $\Delta$. We also display as yellow dots the Bethe roots associated with the exponentially degenerate ground state (``Umklapp'' type excitation), obtained from the former by continuously driving the twist to $\varphi=2\pi$. 
\begin{figure}
\begin{center}
\includegraphics[scale=0.6]{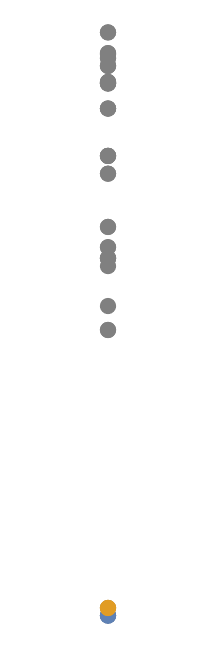}
\hspace{2cm}
\includegraphics[scale=1.2]{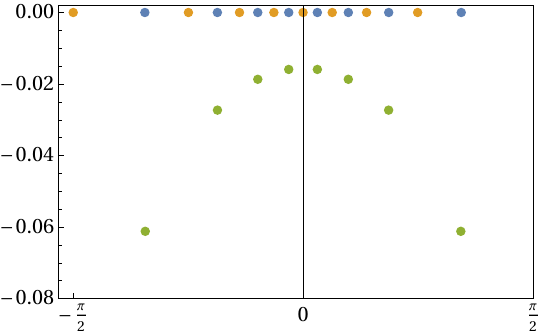}
\caption{(Left) Low-lying spectrum of the XXZ chain obtained by exact diagonalization for $L=16$ sites, $\Delta=2.5$ and zero twist. The colored dots indicate the two exponentially degenerate ground states. (Right) The corresponding distributions of Bethe Roots in the complex plane. The blue and yellow dots correspond to the aforementioned ground states, while the green dots correspond to ground state root distribution for an imaginary twist $\varphi=i$.}  
\label{fig:Betheroots}
\end{center}
\end{figure}
Higher excited states, including those in sectors of nonzero magnetization, are obtained from the ground state by allowing for {\it holes} and {\it strings} in the distribution of Bethe roots, each coming with a finite energy cost \cite{baxter2016exactly,takahashi1999thermodynamics, des1966anisotropic} in agreement with the strong-coupling analysis presented in Section \ref{sec:lowenergy}, together with the spectrum of Figure \ref{fig:Betheroots}, obained by exact diagonalisation.
}

\begin{figure}
\begin{center}
\includegraphics[scale=1]{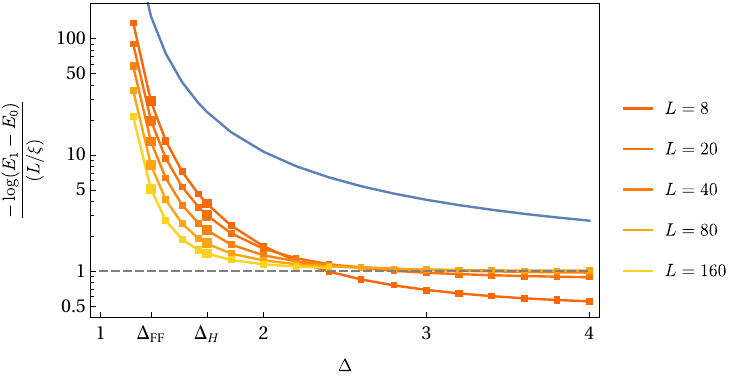}
\caption{The rescaled gap $- \frac{\log(E_1(L)-E_0(L))}{L/\xi}$ between the ground states of the $(0,\varphi_0)$ and $(0,\varphi_1)$ sectors, as a function of $\Delta$. The red to yellow lines show data from Bethe ansatz for increasing system sizes, where we used larger plot marks to evidence the Fib$\times$Fib and Haagerup points. The blue line gives the value of the correlation length $\xi$ as a function of $\Delta$.}  
\label{fig:gap}
\end{center}
\end{figure}

We can now readily use the Bethe ansatz technology to compute the energies $E_0(L)$ and $E_1(L)$ for large values of $L$. For a given value of $L$ and $\eta$, we start by computing the rapidities of the ground state at zero twist by numerically resolving the logarithmic Bethe equations \eqref{eq:logBAE} with a prescribed distribution of the integers $I_j$. From there, we vary the twist continuously, and obtain by a Newton-Raphson method the sets of rapidities for the ground states at twists $\varphi_0$ (imaginary) and $\varphi_1$ (real). Using the above expression for the energies, this gives access to the gap $\delta E(L) = E_1-E_0$. 
Since the expressions of $\varphi_0$ and $\varphi_1$ can be naturally extended to arbitrary $\eta$, we compute the latter for a range of values spanning the gapped phase of the XXZ model, and including the points $\eta_{\text{Fib}\times \text{Fib}}$ and $\eta_{\mathcal{H}_3}$.
Our results are shown on on Figure \ref{fig:gap}, where for large enough $\Delta$ we clearly see the exponential vanishing of the gap, 
\be 
E_1(L)-E_0(L) \sim e^{-L / \xi}  \,.
\label{eq:exponentialdecay}
\ee 
For large enough $\Delta$, the ratio $- \frac{\log(E_1(L)-E_0(L))}{L/\xi}$ indeed quickly converges to $1$ as $L$ grows. As
$\Delta$ decreases to $1$ and as the correlation length grows the convergence is slower, and larger system sizes are
needed to see clearly the exponential decay of the gap.   
 
A mathematically more rigorous derivation of the low lying energies is possible by the method of non-linear integral
equations (NLIE). The idea is to express the Bethe equations as an NLIE and to extract the large volume behaviour
afterwards. For an example of using this method see \cite{nlie-massive-xxz}. We did not perform such computations
adapted to our situation, because both the numerical data and the general arguments appear convincing enough.

\section{Conclusion and Outlook}

We showed that in any fusion category which has a self-dual and non-invertible element, we can construct an integrable anyonic spin chain that is defined from a Temperley-Lieb algebra. This anyonic spin chain is defined by considering the projector on the identity element in the fusion of two of the aforementioned non-invertible elements. Our models are related to so-called Temperley-Lieb interaction models.

Because of the underlying Temperley-Lieb algebra, we were able to access the spectrum of these models by exploiting its irreducible representations and relate them to the spectrum of a twisted XXZ spin chain. We worked out the explicit map for the categories $\mathcal{C} = \mathcal{H}_3$, $\mathcal{C} = \text{Fib}\times \text{Fib}$, and $\mathcal{C} = \mathfrak{psu}(2)_k$ and checked our results with explicit diagonalisation and DMRG.

Because of this map we argue that the anyonic chains that we consider are gapped if the quantum dimension $\delta$ of the non-invertible element is greater than 2. We also show that the correlation length is large when $\delta$ is close to two. This demonstrates that finite size effects can have a large impact while doing numerics. 

There are several directions for further research. In \cite{Corcoran:2024eeh} there was a further conjecture that the anyonic chain defined by the projection on the non-invertible element itself, $p^{(a,a)}$, is gapless and critical. This observation was supported by numerics. However, these models are generically not integrable and  there are currently no analytic tools to address this question. It would be interesting to investigate this further.

It would be interesting to classify the Fusion categories that admit a gapless anyon chain of Temperley-Lieb type. In particular, since the quantum dimensions need to be compatible with the fusion rules, this means that the presence of an element with quantum dimension smaller than 2 will put some restrictions on the allowed fusion rules. It would be further interesting to see which set of CFTs these chains would correspond to in the continuum limit.

In this work we focused on anyonic chains based on Fusion categories with $N^{c}_{ab}\in \{0,1 \}$. It would be interesting to investigate the algebraic structure in models constructed from fusion categories with $N_{ab}^c > 1$. It would also be interesting to study models constructed from fusion 2-categories and beyond.

The models we considered with quantum dimension greater than 2 all come from unitary fusion categories. There exist several non-unitary fusion categories with $-2<\delta \leq 2$, and so the corresponding XXZ chain has $-1<\Delta\leq 1$. One example is the non-unitary counterpart of the fusion category $\mathcal{H}_3$, which has $\delta = \frac{3-\sqrt{13}}{2}$. Given that this lies in the critical range of XXZ, it would be interesting to check if this model itself is critical, or its non-unitary nature obscures the decomposition into TL representations.

Finally, since the models we consider in the paper are integrable, it is reasonable to hope that there are algebraic methods to determine their spectra (and that of general integrable anyonic chains). We leave this as an important direction for future research.

\vspace{1cm}
\noindent {\bf Acknowledgments} 
EV thanks Fabian Essler and Paul Fendley for reminding him of how big the XXZ correlation function can remain at
reasonable distance from the transition point. MdL was supported in part by SFI and the Royal Society for funding under
grants UF160578, RGF$\backslash$ R1$\backslash$ 181011, RGF$\backslash$8EA$\backslash$180167 and RF$\backslash$
ERE$\backslash$ 210373. MdL is also supported by ERC-2022-CoG - FAIM 101088193, which also supports LC. Research at Perimeter Institute is
supported in part by the Government of Canada through the Department of Innovation, Science and Economic Development and
by the Province of Ontario through the Ministry of Colleges and Universities. BP was supported by the NKFIH excellence
grant TKP2021-NKTA-64. 
EV was supported by the grant Emergence-ComplexCité "INTOTHREE".
\bigskip


\appendix

\section{Fusion Categorical Proof of Temperley-Lieb Structure}

\label{sec:appA}
In \ref{par:TLStructure}, it is stated that the operators defined in \eqref{eq:Xdef} satisfy the defining relations of the Temperley-Lieb algebra. An explanation of how the first and third properties in \eqref{eq:TL} follow from the definition of $X_i^{(a)}$ is presented following the statement of the theorem. Here, we present proof that the second property also holds under an appropriate choice of gauge.
\par
As was discussed in \ref{par:TLStructure}, inspection of \eqref{eq:projectorform} provides that $p_i^{(a,1)}=\delta_{x_{i-1},x_{i+1}}p_i^{(a,1)}$. As such, we can define the action of $p_i^{(a,1)}$ by
\begin{equation}
p_{i}^{(a,1)}\ket{x_{i-1}x_ix_{i+1}}=\delta_{x_{i-1}x_{i+1}}\sum_{x_i'\in\mathcal{C}}(F_{x_{i+1}}^{x_{i-1},a,a})_{x_i,1}(F_{x_{i+1}}^{x_{i-1},a,a})_{1,x_i'}^{-1}\ket{x_{i-1}x_i'x_{i+1}}.
\end{equation}
\par
Throughout the following manipulations, the Kronecker delta that was pulled out to the front of the expression for $p_i^{(a,1)}$ is utilised.
\begin{equation}
\begin{aligned}
&p_{i+1}^{(a,1)}p_{i}^{(a,1)}\ket{x_{i-1}x_ix_{i+1}x_{i+2}}\\
&=\delta_{x_{i-1}x_{i+1}}\sum_{x_i'}(F_{x_{i+1}}^{x_{i-1},a,a})_{x_i,1}(F_{x_{i+1}}^{x_{i-1},a,a})_{1,x_i'}^{-1}p_{i+1}^{(a,1)}\ket{x_{i-1}x_i'x_{i+1}x_{i+2}}\\
&=\delta_{x_{i-1}x_{i+1}}\sum_{x_i'}(F_{x_{i+1}}^{x_{i-1},a,a})_{x_i,1}(F_{x_{i+1}}^{x_{i-1},a,a})_{1,x_i'}^{-1}\delta_{x_i'x_{i+2}}\cdots\\
&\hspace{0.5cm}\sum_{x_{i+1}'}(F_{x_{i+2}}^{x_{i}',a,a})_{x_{i+1},1}(F_{x_{i+2}}^{x_{i}',a,a})_{1,x_{i+1}'}^{-1}\ket{x_{i-1}x_i'x_{i+1}'x_{i+2}}\\
&=\delta_{x_{i-1}x_{i+1}}(F_{x_{i+1}}^{x_{i-1},a,a})_{x_i,1}(F_{x_{i+1}}^{x_{i-1},a,a})_{1,x_{i+2}}^{-1}\cdots\\
&\hspace{0.5cm}\sum_{x_{i+1}'}(F_{x_{i+2}}^{x_{i+2},a,a})_{x_{i+1},1}(F_{x_{i+2}}^{x_{i+2},a,a})_{1,x_{i+1}'}^{-1}\ket{x_{i-1}x_{i+2}x_{i+1}'x_{i+2}}\\
\end{aligned}
\end{equation}
\begin{equation}
\begin{aligned}
&\implies p_{i}^{(a,1)}p_{i+1}^{(a,1)}p_{i}^{(a,1)}\ket{x_{i-1}x_ix_{i+1}x_{i+2}}\\
&=\delta_{x_{i-1}x_{i+1}}(F_{x_{i+1}}^{x_{i-1},a,a})_{x_i,1}(F_{x_{i+1}}^{x_{i-1},a,a})_{1,x_{i+2}}^{-1}\sum_{x_{i+1}'\in\mathcal{C}}(F_{x_{i+2}}^{x_{i+2},a,a})_{x_{i+1},1}(F_{x_{i+2}}^{x_{i+2},a,a})_{1,x_{i+1}'}^{-1}\cdots\\ 
&\hspace{0.5cm}\delta_{x_{i-1}x_{i+1}'}\sum_{x_i'\in\mathcal{C}}(F_{x_{i+1}'}^{x_{i-1},a,a})_{x_{i+2},1}(F_{x_{i+1}'}^{x_{i-1},a,a})_{1,x_i'}^{-1}\ket{x_{i-1}x_i'x_{i+1}'x_{i+2}}\\
&=\delta_{x_{i-1}x_{i+1}}(F_{x_{i+1}}^{x_{i-1},a,a})_{x_i,1}(F_{x_{i+1}}^{x_{i-1},a,a})_{1,x_{i+2}}^{-1}(F_{x_{i+2}}^{x_{i+2},a,a})_{x_{i+1},1}(F_{x_{i+2}}^{x_{i+2},a,a})_{1,x_{i+1}}^{-1}\cdots\\
&\hspace{0.5cm}\sum_{x_i'\in\mathcal{C}}(F_{x_{i+1}}^{x_{i-1},a,a})_{x_{i+2},1}(F_{x_{i+1}}^{x_{i-1},a,a})_{1,x_i'}^{-1}\ket{x_{i-1}x_i'x_{i+1}x_{i+2}}\\
&=(F_{x_{i+1}}^{x_{i-1},a,a})_{x_i,1}(F_{x_{i+1}}^{x_{i-1},a,a})_{1,x_{i}}^{-1}(F_{x_{i+2}}^{x_{i+2},a,a})_{x_{i+1},1}(F_{x_{i+2}}^{x_{i+2},a,a})_{1,x_{i+1}}^{-1}p_i^{(a,1)}\ket{x_{i-1}x_ix_{i+1}x_{i+2}}.
\end{aligned}
\end{equation}
Similarly, 
\begin{equation}
\begin{aligned}
&p_{i}^{(a,1)}p_{i-1}^{(a,1)}p_{i}^{(a,1)}\ket{x_{i-2}x_{i-1}x_ix_{i+1}}\\
&=(F_{x_{i+1}}^{x_{i-1},a,a})_{x_i,1}(F_{x_{i+1}}^{x_{i-1},a,a})_{1,x_{i}}^{-1}(F_{x_{i-2}}^{x_{i-2},a,a})_{x_{i+1},1}(F_{x_{i-2}}^{x_{i-2},a,a})_{1,x_{i+1}}^{-1}p_i^{(a,1)}\ket{x_{i-2}x_{i-1}x_ix_{i+1}}.
\end{aligned}
\end{equation}
Defining $X_i^{(a)}=\delta p_i^{(a,1)}$ with
\begin{equation}
\delta = \frac{1}{(F_{x_{i-1}}^{x_{i-1},a,a})_{x_i,1}(F_{x_{i}}^{x_{i},a,a})_{1,x_{i-1}}^{-1}},
\end{equation}
the second Temperley-Lieb property is easily shown to hold. However, in order for $X_i^{(a)}$ to satisfy the first property, $\delta$ must not depend on $x_{i-1}$, $x_i$, or $x_{i+1}$.
\par
To prove that the above choice of $\delta$ is indeed constant, we will use a set of equations derived by manipulating fusion diagrams in \cite{osborne2019fsymbols} (Theorem 5.2 there). There, the equations are derived while working with unitary $F$-symbols and using a different indexing notation, but in the conventions used here, the equations are:
\begin{equation}
(F_d^{\alpha,b,c})_{e,f}^{-1}(F_{f}^{g,h,b})_{\alpha,i}=\sum_{j\in \mathcal{C}}(F_d^{g,h,e})_{\alpha,j}(F_j^{h,b,c})_{e,i}^{-1}(F_d^{g,i,c})_{j,f}^{-1}.
\label{eq:InvPent}
\end{equation}
Setting $b=c=h=a$, $e=i=1$, $g=f=x_{i-1}$, and $\alpha=d=x_i$, the only non-zero term on the right-hand side has $j=a$:
\begin{equation}
(F_{x_{i}}^{x_{i},a,a})_{1,x_{i-1}}^{-1}(F_{x_{i-1}}^{x_{i-1},a,a})_{x_i,1}=(F_{x_i}^{x_{i-1},a,1})_{x_i,a}(F_a^{a,a,a})_{1,1}^{-1}(F_{x_i}^{x_{i-1},1,a})_{a,x_{i-1}}^{-1}.
\end{equation}
It was proved in \cite{gert_thesis} (Proposition 4.1.3) that for any fusion category, a gauge choice can be made such that the $F$-symbols with the identity object as an upper inner index are all 1. Choosing this gauge gives us
\begin{equation}
(F_{x_{i}}^{x_{i},a,a})_{1,x_{i-1}}^{-1}(F_{x_{i-1}}^{x_{i-1},a,a})_{x_i,1}=(F_a^{a,a,a})_{1,1}^{-1}.
\label{eq:InvPentSpecial}
\end{equation}
Thus, we have that
\begin{equation}
X_i^{(a)}=\frac{1}{(F_a^{a,a,a})_{1,1}^{-1}}p_i^{(a,1)}
\end{equation}
is a Temperley-Lieb operator .
\par
In the special case of unitary fusion categories, $\frac{1}{(F_a^{a,a,a})_{1,1}}=\frac{1}{(F_a^{a,a,a})_{1,1}^{-1}}=\kappa_a d_{FP,a}$, where $\kappa_a=\pm 1$ is the `Frobenius-Schur indicator' for the particle species $a$ and $d_{FP,a}$ is its Frobenius Perron dimension - this results in Temperley-Lieb parameter $\delta=\kappa_ad_{FP,a}$.

\section{Equivalence to TLI and ADE Models}

\label{sec:appB}

The TLI models' Temperley-Lieb operator \cite{Pearce:1990ila,Pasquier:1987xj,PASQUIER1987162} is defined as
\begin{equation}
\tilde{e}_i\ket{x_{i-1}x_ix_{i+1}}=\delta_{x_{i-1}x_{i+1}}\sum_{x_i'}\frac{(S_{x_i'}S_{x_i})^{1/2}}{S_{x_{i-1}}}\ket{x_{i-1}x_i'x_{i+1}}
\end{equation}
(leaving out a factor of $\beta^{-1}$ in Pasquier's definition to align with our definition of Temperley-Lieb operators), where $S_{x_i}$ are defined as the $x_i$ entry of the eigenvector of the adjacency matrix with largest eigenvalue.
\par
If object/particle $a$ is on the external edges of our `anyonic chain', the adjacency matrix can be written in terms of fusion coefficients as
\begin{equation}
C_a=\sum_{b}\sum_c N_c^{b,a}e_{b,c}.
\end{equation}
The eigen-equation is
\begin{equation}\
\left(C_a\vec{v}\right)_c=\sum_{d}N_d^{c,a}\vec{v}_d=\lambda_a\vec{v}_c.
\end{equation}
Recalling that for unitary fusion categories, the quantum dimensions are a `solution to the fusion rules' in the sense that
\begin{equation}
\text{dim}\hspace{0.05cm}a\hspace{0.05cm}\text{dim}\hspace{0.05cm}b=\sum_{c}N_{a}^{a,b}\text{dim}\hspace{0.05cm}c,
\end{equation}
the eigen-equation of the adjacency matrix is solved by $\lambda_a=\text{dim}\hspace{0.05cm}a, \space \vec{v}_b=\text{dim}\hspace{0.05cm}b$. There may be multiple solutions to the quantum dimension formula; the TLI construction of ADE models uses the Frobenius-Perron eigenvector/eigenvalue. Therefore the TLI operator on an anyonic chain with external edges of type $a$ has Temperley-Lieb parameter $d_{FP,a}$ and is given explicitly by
\begin{equation}
\tilde{e}_i\ket{x_{i-1}x_ix_{i+1}}=\delta_{x_{i-1}x_{i+1}}\sum_{x_i'}\frac{(d_{FP,x_i'}d_{FP,x_i})^{1/2}}{d_{FP,x_{i-1}}}\ket{x_{i-1}x_i'x_{i+1}}.
\end{equation}
The Frobenius-Perron dimension of object $a$ may only coincide with the quantum dimension of the objects if the fusion category is unitary. Therefore, if our Temperley-Lieb operator is defined using a non-unitary fusion category, it is guaranteed not to be an operator defined by TLI construction of ADE models. The relationship between our models with different Temperley-Lieb parameters and the TLI/ADE model construction is discussed in \ref{par:TLI/ADE}.
\par
Our Temperley-Lieb operator is defined by
\begin{equation}
X_i^{(a)}\ket{x_{i-1}x_ix_{i+1}}=\delta_{x_{i-1}x_{i+1}}\sum_{x_i'}\frac{(F_{x_{i+1}}^{x_{i-1},a,a})_{x_{i},1}(F_{x_{i+1}}^{x_{i-1},a,a})_{1,x_i'}^{-1}}{(F_{a}^{a,a,a})_{1,1}^{-1}}\ket{x_{i-1}x_i'x_{i+1}}.
\end{equation}
Therefore, asking whether our Temperley-Lieb operator is equivalent (under some choice of gauge for our $F$-symbols) to a TLI operator reduces asking whether a solution exists to the equation
\begin{equation}
\frac{g_{x_{i-1}ax_i}g_{x_iax_{i+1}}}{g_{x_{i-1}ax_i'}g_{x_i'ax_{i+1}}}\frac{(F_{x_{i+1}}^{x_{i-1},a,a})_{x_{i},1}(F_{x_{i+1}}^{x_{i-1},a,a})_{1,x_i'}^{-1}}{(F_{a}^{a,a,a})_{1,1}^{-1}}=\frac{(d_{FP,x_i'}d_{FP,x_i})^{1/2}}{d_{FP,x_{i-1}}},
\label{eq:1stCondition}
\end{equation}
while keeping the F-symbols in a gauge where F-symbols with the vacuum as an upper-inner index are 1.
\par
Noticing the symmetry of the right-hand side of this equation, we have that any solution must have
\begin{equation}
    \frac{g_{x_{i+1}ax_i}g_{x_iax_{i+1}}}{g_{x_{i+1}ax'_i}g_{x'_iax_{i+1}}}=\sqrt{\frac{(F_{x_{i+1}}^{x_{i+1,a,a}})_{1,x_i}^{-1}(F_{x_{i+1}}^{x_{i+1,a,a}})_{x'_i,1}}{(F_{x_{i+1}}^{x_{i+1,a,a}})_{1,x'_i}^{-1}(F_{x_{i+1}}^{x_{i+1,a,a}})_{x_i,1}}}
\end{equation}
this choice of gauge leaves the condition for equivalence between our models and TLI models as
\begin{equation}
    \frac{\sqrt{(F_{x_{i+1}}^{x_{i+1},a,a})_{1,x_i}^{-1}(F_{x_{i+1}}^{x_{i+1},a,a})_{x'_i,1}{(F_{x_{i+1}}^{x_{i+1},a,a})_{1,x'_i}^{-1}(F_{x_{i+1}}^{x_{i+1},a,a})_{x_i,1}}}}{(F_{a}^{a,a,a})_{1,1}^{-1}}=\frac{\sqrt{d_{FP,x_i'}d_{FP,x_i}}}{d_{FP,x_{i-1}}}.
\end{equation}
Now, reusing \eqref{eq:InvPentSpecial}, the condition reduces to 
\begin{equation}
    \sqrt{\frac{(F_{x_{i+1}}^{x_{i+1},a,a})_{x'_i,1}(F_{x_{i+1}}^{x_{i+1},a,a})_{x_i,1}}{(F_{x_{i}}^{x_i,a,a})_{x_{i+1},1}(F_{x_{i}'}^{x_i',a,a})_{x_{i+1},1}}}=\frac{\sqrt{d_{FP,x_i'}d_{FP,x_i}}}{d_{FP,x_{i-1}}}.
    \label{eq:FinalCond}
\end{equation}
The following relation was derived for unitary fusion categories in \cite{Wolf:2020qdo} (6.19):
\begin{equation}
    |(F_a^{a,b,\bar{b}})_{c,1}|^2=\frac{\text{dim}\hspace{0.05cm}c}{\text{dim}\hspace{0.05cm}a\hspace{0.05cm}\text{dim}\hspace{0.05cm}b}.
\end{equation}
This relation implies that \eqref{eq:FinalCond} is always satisfied. Therefore, when we define our operators using unitary fusion categories such that $\kappa_a=1$, there is always some gauge such that our Temperley-Lieb operators have the same matrix elements as one produced via the TLI construction on the same set of fusion rules.
\par
In fact, the equivalence of our models (when defined using unitary fusion categories such that $\kappa_a=1$) to a model produced from the TLI construction holds at the level of energy spectra in any given choice of gauge. To see this, notice that the matrix elements of our local Temperley-Lieb operators in the $\{\ket{x_{i-1}x_ix_{i+1}}\}$ basis change under gauge transformations as given on the left hand side of \eqref{eq:1stCondition}. However, the matrix elements of the operator after the gauge transformation in the $\{\frac{1}{g_{x_{i-1}ax_i}g_{x_iax_{i+1}}}\ket{x_{i-1}x_ix_{i+1}}\}$ basis are equal to the matrix elements of the untransformed operator in the original basis, so the operator's eigenvalues are unaffected. This logic can easily be extended to the total Hamiltonian's matrix elements, showing that the energy spectrum is unaffected by the gauge transformations of $F$-symbols.
\par
Therefore, whenever we define our Temperley-Lieb operator using a unitary fusion category such that $\kappa_a=1$, our model has the same energy spectrum as a model produced via the TLI construction on the same set of fusion rules.

\section{Equivalence of Coupled Golden Chains to a Fib$\times$Fib Chain}

\label{sec:appC}

Considering systems A and B to be golden chains made up of particles $\{1,2\}$ and $\{1,3\}$ respectively and the fusion rules of Fib $\times$ Fib \cite{vercleyen_slingerland_anyonwiki_2025}, we have that a pair of Fibonacci anyonic chains is isomorphic to a Fib$\times$Fib anyonic chain with its external edges being object 4. The isomorphism between pairs of edges in systems A and B, and single Fib$\times$Fib edges in system AB is presented in the following table:
\begin{center}
\begin{tabular}{c | c c }
 System AB & System A & System B\\
 \hline
 1 & 1 & 1\\
 2 & 2 & 1\\
 3 & 1 & 3\\
 4 & 2 & 3 
\end{tabular}
\end{center}
This isomorphism allows us to generate a subset of the set of possible Fib$\times$Fib $F$-symbols from two sets of Fib $F$-symbols. This follows from the fact that $F$-symbols are solutions to the pentagon equations for a given set of fusion rules - algebraic expressions of the commutation diagram \eqref{eq:pentdiagram}.\\
Knowing that any Fib$\times$Fib fusion diagram is isomorphic to a pair of Fib fusion diagrams, two sets of solutions to the Fib pentagon equations can be used to solve the Fib$\times$Fib pentagon equations. Specifically, each Fib$\times$Fib $F$-symbol may simply be written as a product of the Fib $F$-symbols with indices which correspond via the isomorphism.\\
The interaction term in equation (2.1) of \cite{Antunes:2025huk} is
\begin{equation}
H_{\text{int}}=-K\sum_i d_2 d_3 p_{i, \text{Fib}}^{(a,1)}\otimes p_{i, \text{Fib}}^{(a,1)}=-\sum_i d_4 p_{i, \text{Fib}}^{(a,1)}\otimes p_{i, \text{Fib}}^{(a,1)}.
\end{equation}
Therefore, to show that this Hamiltonian is (up to the factor of $K$) our Temperley-Lieb Hamiltonian in the picture of a single Fib$\times$Fib chain, 
\begin{equation}
H_{\text{TL}}=-d_4\sum_i p_{i, \text{Fib}\times\text{Fib}}^{(a,1)},
\end{equation}
we just need to show that
\begin{equation}
(F_{\tilde{x}_{i-1}}^{\tilde{x}_{i+1},2,2})_{\tilde{x}_i,1}(F_{\tilde{x}_{i-1}}^{\tilde{x}_{i+1},2,2})_{\tilde{x}_i',1}^\dagger (F_{\bar{x}_{i-1}}^{\bar{x}_{i+1},3,3})_{\bar{x}_i,1}(F_{\bar{x}_{i-1}}^{\bar{x}_{i+1},3,3})_{\bar{x}_i',1}^\dagger=(F_{x_{i-1}}^{x_{i+1},4,4})_{x_i,1}(F_{x_{i-1}}^{x_{i+1},4,4})_{x_i',1}^\dagger,
\label{eq:FibFibCond}
\end{equation}
where $x_j=\tilde{x}_j\bar{x}_j$.\\
This equation is not gauge invariant, so it does not hold for all sets of Fib$\times$Fib $F$-symbols and cannot be proved from the pentagon equations. However, if the Fib$\times$Fib gauge variables are fixed to be products of the corresponding two Fib gauge variables via the isomorphism established above, the remaining gauge freedom doesn't affect this equation. Further, \eqref{eq:FibFibCond} follows directly from the method of generating Fib$\times$Fib $F$-symbols from two sets of Fib $F$-symbols, whereby
\begin{equation}
(F_{x_{i-1}}^{x_{i+1},4,4})_{x_i,1}=(F_{\tilde{x}_{i-1}}^{\tilde{x}_{i+1},2,2})_{\tilde{x}_i,1}(F_{\bar{x}_{i-1}}^{\bar{x}_{i+1},3,3})_{\bar{x}_i,1}.
\end{equation}
Therefore, if the Fib$\times$Fib $F$-symbols that are being used are generated in this way, $H_{\text{TL}}$ and $H_{\text{int}}$ have the same matrix elements.

\addcontentsline{toc}{section}{References}
\bibliography{GeneralFusionTL}
\bibliographystyle{utphys}

\end{document}